\PassOptionsToPackage{colorlinks=true,linkcolor=blue,citecolor=blue,urlcolor=blue}{hyperref}
\documentclass[aps, superscriptaddress, floatfix, notitlepage, twocolumn, nofootinbib]{revtex4-2}
\usepackage[utf8]{inputenc}
\usepackage{amsfonts}
\usepackage{amssymb}
\usepackage{amsmath}
\usepackage{amsfonts}
\usepackage{bbm}
\usepackage[normalem]{ulem}
\usepackage{graphicx}
\usepackage{dcolumn}
\usepackage{bm}
\usepackage{amsmath}
\usepackage{orcidlink}
\usepackage{textcase}
\usepackage[T1]{fontenc}
\usepackage[utf8]{inputenc}
\usepackage{dsfont}
\usepackage{graphicx}
\usepackage{dcolumn}
\usepackage[colorlinks=true,linkcolor=blue,citecolor=blue,urlcolor=blue]{hyperref}
\begin{document}

\preprint{APS/123-QED}

\newcommand{\upv}{Fisika Aplikatua Saila, Gipuzkoako Ingeniaritza Eskola, University of the Basque Country (UPV/EHU), Europa Plaza 1, 20018 Donostia/San Sebastián, Spain}
\newcommand{\cfm}{Centro de Física de Materiales (CFM-MPC), CSIC-UPV/EHU, Manuel de Lardizabal Pasealekua 5, 20018 Donostia/San Sebastián, Spain}
\newcommand{\dipc}{Donostia International Physics Center (DIPC), Donostia, 20018, Spain}
\newcommand{\INMA}{Instituto de Nanociencia y Materiales de Aragón (INMA), CSIC-Universidad de Zaragoza, Zaragoza,
50009, Spain}

\title{First-order correction to the self-consistent harmonic approximation free energy: Impact on the negative thermal expansion and phonon dynamics in ScF$_3$}

\author{Jincheng Yue\orcidlink{0000-0001-5107-4139}}
\email{jyue001@ikasle.ehu.eus}
\affiliation{\cfm}
\affiliation{\upv}

\author{Manex Alkorta\orcidlink{0009-0003-8109-9216}}
\affiliation{\cfm}
\affiliation{\upv}

\author{\DJ{}or\dj{}e Dangi\'{c}\orcidlink{0000-0002-4639-184X}}
\affiliation{\cfm}
\affiliation{\upv}

\author{Yue-Wen Fang\orcidlink{0000-0003-3674-7352}}
\email{yuewen.fang@csic.es}
\affiliation{\INMA}
\affiliation{\cfm}

\author{Ion Errea\orcidlink{0000-0002-5719-6580}}
\email{ion.errea@ehu.eus}
\affiliation{\cfm}
\affiliation{\upv}
\affiliation{\dipc}

\begin{abstract}
Strongly anharmonic crystals are often treated within the self-consistent harmonic approximation, which employs variationally optimized Gaussian density matrices, and the related self-consistent phonon theory.
These approaches capture substantial phonon renormalization but exclude connected non-Gaussian correlations in the free energy.
Here, we derive the first-order post-variational correction generated by the lowest non-vanishing normal-ordered residual vertex. Normal ordering the residual interaction around the optimized state shows that the linear and quadratic vertices vanish at the stationary variational solution, leaving a fluctuation-dressed cubic interaction whose second-order vacuum and two-point contractions generate the free-energy correction and the dynamical bubble self-energy, respectively. In ScF$_3$, the variational optimization on the full Born-Oppenheimer potential accounts for most of the shift in the thermal-expansion crossover temperature between negative and positive values relative to the result obtained with self-consistent phonon theory when the potential is truncated at fourth order. The first-order connected correction changes the crossover only modestly, but its volume derivative appreciably shifts the pressure-remapped cubic local-stability limit. Using the bare cubic vertex appears to improve agreement with the thermodynamic measurements, yet it systematically underestimates the measured zone-boundary tilt-mode energies at high temperatures. The fluctuation-dressed interaction gives more moderate thermodynamic corrections but yields phonon energies in better agreement with experiments. These results show that connected non-Gaussian correlations can significantly affect free-energy derivatives and soft-mode dynamics near a structural instability, and identify the optimized Gaussian state as the natural reference for constructing residual interactions beyond the stochastic self-consistent harmonic approximation.
\end{abstract}

\maketitle
\section{INTRODUCTION}
\par Strong lattice anharmonicity governs the structural and thermodynamic behavior of many crystalline materials, particularly those characterized by shallow Born-Oppenheimer (BO) potential-energy landscapes, and large-amplitude ionic quantum and thermal fluctuations~\cite{monacelli2021stochastic, tadano2022first, Zhao2025IonicAnharmonicity,PRL2026-STO-4293-1g2r,HaoGAO2026_H3S,alkorta2026symmetry}. In this regime, a perturbative treatment built on a harmonic reference may converge slowly or fail altogether. This difficulty is especially severe near structural instabilities, where the relevant free-energy differences are small and the equilibrium structure is determined by the anharmonic part of the potential~\cite{alkorta2026symmetry,errea2016quantum, xia2020microscopic, zheng2022anharmonicity, zheng2024unravelling, dangic2025lattice,yue2025interlayer}. Predicting thermal expansion, phase stability, and temperature-dependent lattice dynamics therefore requires methods that go beyond low-order perturbation theory. 

\par To this end, several approaches have been developed in the latest decades. Ab initio molecular dynamics (AIMD)~\cite{balbuena1999molecular} samples the BO potential-energy surface without imposing an explicit low-order Taylor expansion, but treats the nuclei classically using Newtonian dynamics. Path-integral molecular dynamics (PIMD)~\cite{marx1996ab, tuckerman1996efficient} incorporates nuclear quantum fluctuations and can, in principle, provide an essentially exact statistical treatment of the ionic degrees of freedom. However, this accuracy comes at a substantial computational cost because PIMD requires
sampling an enlarged path-integral representation of the nuclear degrees
of freedom, often over a large number of configurations. Effective-harmonic and related variational approaches, including Gaussian variational methods~\cite{hooton1955li, koehler1966theory,errea2011anharmonic}, self-consistent phonon (SCP) theories~\cite{klein1972rise}, and temperature-dependent effective-potential (TDEP) methods~\cite{hellman2011lattice}, offer a more direct route to temperature-renormalized phonons and approximate anharmonic free energies. Their accuracy, however, depends on the choice of reference system, the definition of the effective interactions, and the treatment of residual anharmonic contributions outside the chosen approximation.

\par Within this broader class of Gaussian variational approaches, the stochastic self-consistent harmonic approximation (SSCHA) ~\cite{errea2014anharmonic,bianco2017second,monacelli2018pressure,monacelli2021stochastic} provides a powerful framework for strongly anharmonic solids by variationally optimizing a Gaussian ionic density matrix with respect to both the average atomic positions and an auxiliary force-constant matrix, while sampling the complete BO potential without requiring a finite-order Taylor expansion. The resulting centroids, auxiliary phonons, propagators, and SSCHA variational free energy incorporate thermal and quantum fluctuations nonperturbatively. The same framework yields the positional free-energy Hessian governing local structural stability considering anharmonicity~\cite{bianco2017second}, offering a formidable tool to characterize both first- and second-order displacive phase transitions, and admits a time-dependent extension for anharmonic spectral functions and phonon quasiparticles~\cite{monacelli2021time, siciliano2023wigner,lihm_gaussian_2021}. Nevertheless, optimization within the Gaussian manifold does not make the SSCHA variational free energy identical to the exact free energy~\cite{monacelli2025analyzing}. Once the optimal Gaussian reference has been determined, the exact Hamiltonian may be decomposed into the optimized auxiliary Hamiltonian and a residual interaction, and the exact free energy expressed as a connected cumulant expansion around that reference. The SSCHA variational free energy contains the free energy of the optimized Gaussian reference together with the first cumulant of the residual interaction. Higher connected cumulants recover residual correlations not included in the Gaussian variational functional. Because the first cumulant of the centered residual interaction vanishes, the second cumulant constitutes the leading correction beyond the SSCHA while retaining its nonperturbatively optimized Gaussian reference. Such corrections can become particularly important when observables are controlled by derivatives of small free-energy variations. 

\par Scandium trifluoride offers a stringent example~\cite{li2011structural}. Its open cubic framework hosts strongly anharmonic zone-boundary octahedral-tilt modes~\cite{handunkanda2015large,wei2020colossal}, exhibits pronounced negative thermal expansion over a broad temperature range~\cite{greve2010pronounced,van2016anomalous,dove2020quantitative,jacs2016-ScF3-NTE}, and undergoes a pressure-induced transition from cubic \(Pm\bar{3}m\) to rhombohedral \(R\bar{3}c\)~\cite{aleksandrov2009pressure, greve2010pronounced}. Recent self-consistent phonon calculations~\cite{oba2019first} have shown that contributions beyond the variational free-energy can indeed substantially alter the predicted thermal expansion of ScF\(_3\). However, conventional SCP implementations generally employ BO potentials truncated at finite anharmonic orders~\cite{tadano2015self, tadano2018first}. For a quartic-truncated Hamiltonian, usually the case, this implies that the anharmonic interaction vertices reduce to the bare third- and fourth-order vertices defined from the derivatives of BO potential. In the SSCHA method, by contrast, the optimized Gaussian state samples the complete BO surface, and fluctuations of all orders dress the interacting vertices. Extending beyond-SSCHA corrections to this setting therefore requires not only evaluating an additional connected free-energy contribution, but also defining the residual cubic interaction consistently with the fully optimized Gaussian reference.

\par In this work, we formulate a finite-temperature post-variational expansion around the optimized SSCHA Gaussian reference by normal ordering the residual interaction with respect to the corresponding Gaussian density matrix in order to define a consistent expansion of the SSCHA variational free energy. At lowest order, the expansion yields the cubic vacuum-bubble free-energy correction with averaged third-order anharmonic vertices and not those obtained from derivatives~\cite{oba2019first}.  
Applied to ScF\(_3\), the SSCHA variational minimization considering the full BO potential accounts for most of the shift of the negative thermal-expansion temperature range relative to the SCP result that truncates the expansion of the BO potential at quartic order. The connected cubic correction to the SSCHA free energy results in only a slight change in the temperature crossover from negative to positive thermal expansion, but its volume derivative appreciably modifies the volume-to-pressure mapping of the cubic stability limit. A diagnostic calculation using the bare BO cubic vertices produces larger thermodynamic shifts in the direction of experiment but oversoftens the measured zone-boundary tilt modes at \(300~\mathrm{K}\), whereas the Gaussian-averaged vertices yield  temperature-dependent phonon frequencies in agreement with experiments. These results demonstrate that average vertices can significantly affect thermodynamic derivatives and soft-mode dynamics near a structural instability, and establish fluctuation-dressed  interactions as the natural basis for expansions beyond the SSCHA optimized Gaussian variational state.

\par The remainder of the paper is organized as follows. In Sec.~\ref{sec:sscha} we review the SSCHA variational formalism and introduce the free-energy Hessian and dynamical SSCHA theory used throughout this work. In Sec.~\ref{sec:cubbic} we derive the lowest-rank non-vanishing correction beyond the SSCHA variational free energy and discuss the consistent construction of the residual interaction around the optimized Gaussian state. In Sec.~\ref{app:sscha_gruneisen} we derive the mode-resolved Gr\"uneisen
parameters and their relation to thermal expansion
within the variational SSCHA framework. Sec.~\ref{sec:methods} describes the computational details used for the SSCHA and SCP calculations for ScF$_3$. In Sec.~\ref{sec:results} we present the results for the thermal expansion, the pressure–temperature evolution of the stability limit of the cubic phase, the pressure-induced structural instability, and the anharmonic phonon spectra of ScF$_3$. Finally, Sec.~\ref{sec:conclusions} summarizes our main conclusions. The manuscript is completed with several appendices.

\section{THE STOCHASTIC SELF-CONSISTENT HARMONIC APPROXIMATION}
\label{sec:sscha}
\subsection{Variational formulation of the SSCHA}
\par The stochastic self-consistent harmonic approximation~\cite{errea2014anharmonic,bianco2017second,monacelli2018pressure,monacelli2021stochastic} is a variational method used to determine the structural and vibrational properties of materials at finite temperatures, taking into account at the same time the quantum nature of ions. For fixed temperature $T$ and cell vectors $\left \{ \vec{a}_i  \right \} $, the variational principle minimizes the following free-energy functional with a trial density matrix $\tilde{\rho}_{\boldsymbol{\mathcal{R}},\mathbf{\Phi}}$ with respect to the centroid positions $\boldsymbol{\mathcal{R}}$ and auxiliary force constants (FCs) $\mathbf{\Phi}$ (bold symbols refer to vectors and tensors in compact notation),
\begin{equation}
\mathcal{F} (\boldsymbol{\mathcal{R}} ,\mathbf{\Phi}, \left \{ \vec{a}_i  \right \}   )= \left \langle K+V_{\mathrm{BO}}(\boldsymbol{R}, \left \{ \vec{a}_i  \right \}  ) \right \rangle _{\tilde{\rho}_{\boldsymbol{\mathcal{R}} ,\mathbf{\Phi} }} - TS_{\mathrm{ion}}[\tilde{\rho}_{\boldsymbol{\mathcal{R}} ,\mathbf{\Phi} }].
\end{equation}
Here $K$ is kinetic energy, $V_{\mathrm{BO}}(\boldsymbol{R},\left \{ \vec{a}_i  \right \} )$ is the BO potential. $S_{\mathrm{ion}}$ is the ionic entropy defined as $S_{\rm ion}[\tilde{\rho}_{\boldsymbol{\mathcal R},\boldsymbol\Phi}]=-k_B\mathrm{Tr}[\tilde{\rho}_{\boldsymbol{\mathcal R},\boldsymbol\Phi}\ln\tilde{\rho}_{\boldsymbol{\mathcal R},\boldsymbol\Phi}]$. \(\langle\cdots\rangle_{ \tilde{\rho}_{\boldsymbol{\mathcal R},\mathbf{\Phi}}}\) denotes the quantum thermal average over the  Gaussian density matrix. After the minimization, the equilibrium SSCHA free energy,
\begin{equation}
F_{\tilde{\rho}_{\boldsymbol{\mathcal{R}} ,\mathbf{\Phi}}}(\left \{ \vec{a}_i  \right \} )= \underset{\boldsymbol{\mathcal{R}},\mathbf{\Phi}}{\mathrm{min}}   \mathcal{F}(\boldsymbol{\mathcal{R}},\mathbf{\Phi}, \left \{ \vec{a}_i  \right \} )=\mathcal{F}(\boldsymbol{\mathcal{R}}_{\mathrm{eq} },\mathbf{\Phi_{\mathrm{eq} }}, \left \{ \vec{a}_i  \right \} ),
\end{equation}
is given in terms of the equilibrium configuration $\boldsymbol{\mathcal{R}}_{\mathrm{eq} }$ and auxiliary FCs $\mathbf{\Phi_{\mathrm{eq}}}$.

\par Within the SSCHA framework, the central approximation is to constrain the probability distribution associated with the trial density matrix to be Gaussian. Therefore, we can define the auxiliary harmonic Hamiltonian $\mathcal{H}_{\boldsymbol{\mathcal{R}} ,\mathbf{\Phi}}$ as 
\begin{equation}
\mathcal{H}_{\boldsymbol{\mathcal{R}} ,\mathbf{\Phi}} = K +
\mathcal{V}_{\boldsymbol{\mathcal{R}} ,\mathbf{\Phi}} ,
\label{H}
\end{equation}
with
\begin{equation}
\mathcal{V}_{\boldsymbol{\mathcal{R}} ,\mathbf{\Phi}}=  \frac{1}{2} \sum_{ab} (R_a-\mathcal{R}_a){\Phi}_{ab}(R_b-\mathcal{R}_b), 
\end{equation}
where $R$ denotes the ionic position operator, and all indices $a$, $b$, etc. run over both atomic and Cartesian coordinates from 1 to 3$N_a$, with $N_a$ the number of atoms in the supercell used in the calculations. Introducing the auxiliary harmonic Hamiltonian, the free-energy can be rewritten as
\begin{equation}
F_{\tilde{\rho}_{\boldsymbol{\mathcal{R}} ,\mathbf{\Phi}}}(\left \{ \vec{a}_i  \right \} )=\underset{\boldsymbol{\mathcal{R}},\mathbf{\Phi}}{\mathrm{min}} \left[ F_{\mathbf{\Phi}}+\left \langle V_{\mathrm{BO}}- \mathcal{V}_{\boldsymbol{\mathcal{R}} ,\mathbf{\Phi}}\right \rangle _{\tilde{\rho}_{\boldsymbol{\mathcal{R}} ,\mathbf{\Phi} }} \right], \label{F_F}
\end{equation}
where $F_{\mathbf{\Phi}}$ is the auxiliary harmonic free energy:
\begin{equation}
F_{\mathbf{\Phi}}(T)=\sum_{\mu =1}^{3N_a} \left [ \frac{\hbar \omega _\mu}{2} +\frac{1}{\beta }\mathrm{ln}(1-e^{-\beta\hbar \omega _\mu})   \right ].  \label{H_F}
\end{equation}
Here $\omega_\mu^2$ are the eigenvalues of the mass-weighted
auxiliary force-constant matrix
$\Phi_{ab}/\sqrt{M_aM_b}$, and $\omega_\mu$ are the corresponding
auxiliary phonon  frequencies.

\subsection{Free-energy Hessian}
\par The Hessian of the positional free-energy, defined as the mass-weighted matrix of second derivatives with respect to atomic coordinates, $D_{ab}^{F }=1/\sqrt{M_aM_b} \partial^2F/\partial \mathcal{R}^a\mathcal{R}^b$, determine the dynamical stability of a structure consistently accounting for both thermal and quantum anharmonic effects. We can represent the positional free-energy Hessian matrix in a compact tensor notation~\cite{bianco2017second,monacelli2021stochastic},
\begin{equation}
\boldsymbol{D}^{F} = \boldsymbol{D}^{S }+\mathbf{\Pi} (0), \label{Full}
\end{equation}
where
\begin{equation}
\mathbf{\Pi} (0)= \boldsymbol{D}^{(3)} : \mathbf{\Lambda}(0):\left [ \mathbbm{1}-\boldsymbol{D}^{(4)} :\mathbf{\Lambda}(0)\right ]^{-1}:\boldsymbol{D}^{(3)}.
\end{equation}
Here
$D^{S}_{ab}=\Phi_{ab}/\sqrt{M_aM_b}$ is the SSCHA auxiliary
dynamical matrix. The tensors $D^{(3)}$ and $D^{(4)}$ are the
corresponding mass-weighted third- and fourth-order
Gaussian-averaged vertices, respectively, and $\Lambda(0)$ is the
static two-phonon propagator. The colon denotes contraction over two
tensor indices.

\par Within the bubble approximation, neglecting the contribution of higher-order terms in the geometric series expansion, the free-energy Hessian matrix becomes
\begin{equation}
\boldsymbol{D}^{F} = \boldsymbol{D}^{S }+\mathbf{\Pi}^{(\mathrm{B})} (0), \label{Bubble}
\end{equation}
where 
\begin{equation}
\mathbf{\Pi}^{(\rm B)} (0)= \boldsymbol{D}^{(3)} : \mathbf{\Lambda}(0):\boldsymbol{D}^{(3)}
\label{s-bubble}
\end{equation}
is the static bubble self-energy.

\subsection{Dynamical theory and spectral functions}
\par Within the time dependent self-consistent harmonic approximation (TD-SCHA)~\cite{monacelli2021time, siciliano2023wigner}, it is possible to define a dynamical extension for the one-phonon Green function $\boldsymbol{G}(\mathbf{q},z)$ in momentum space, 
\begin{equation}
\boldsymbol{G}^{-1}(\mathbf{q},z)=z^{2}\mathbbm{1}-(\boldsymbol{D}^{S}(\mathbf{q})+\boldsymbol{\Pi}(\mathbf{q},z)), 
\label{z-G}
\end{equation}
where $\boldsymbol\Pi(\mathbf{q},z)$ is the TD-SCHA self-energy with momentum $\mathbf{q}$.
In the static approximation, the self-energy in Eq.~(\ref{z-G}) reduces to Eq.~(\ref{s-bubble}) by setting ($z=0$). To calculate the phonon spectra, we retain the frequency dependence of the bubble self-energy, ${\mathbf{\Pi}} (\mathbf{q},z) \approx \mathbf{\Pi}^{(\mathrm{B})}(\mathbf{q},z)$. The retarded Green function is then obtained by evaluating Eq.~(\ref{z-G}) at $(z=\Omega+i0^+)$, and its imaginary part gives the spectral function,
\begin{equation}
\begin{aligned}
\sigma(\mathbf{q},\Omega)
= &-\frac{\Omega}{\pi}\,\mathrm{Im} \mathrm{Tr} 
\Big[
(\Omega + \mathrm{i}\,\delta)^2 \mathbbm{1}
 \\
&
-\Big(
\boldsymbol{D}^{S}(\mathbf{q})
+ \boldsymbol{\Pi}^{(\mathrm{B})}(\mathbf{q},\Omega+\mathrm{i}\delta)
\Big)
\Big]^{-1},
\end{aligned}
\end{equation}
with $\delta$ arbitrary small, but finite, positive smearing
value. 
Since the dynamical bubble self-energy is naturally formulated in the phonon eigenmode basis, its matrix elements are denoted by $\mathbf{\Pi}^{(\mathrm{B})}_{\mu \nu }(\mathbf{q},\omega  )$, where $\mu$ and $\nu$ label the vibrational branches at a given $\mathbf{q}$. 
In general, the self-energy is not diagonal in mode space, and the off-diagonal components $\mathbf{\Pi}^{(\mathrm{B})}_{\mu \nu }(\mathbf{q},\omega  ) $ $(\mu \ne \nu)$ account for hybridization between different phonon branches. 

\par In this work, to track the spectral function of the phonon modes we adopt the no mode-mixing approximation, retaining only the diagonal component $\mathbf{\Pi}^{(\mathrm{B})}_{\mu \mu }(\mathbf{q},\omega  )$. Within this approximation, the mode-resolved spectral function can be expressed as
\begin{equation}
\begin{aligned}
\sigma_{\mu}(\mathbf{q},\Omega)
= \frac{1}{2} \Bigg[
& \frac{1}{\pi}
\frac{-\mathrm{Im}\,\mathcal{Z}_{\mu}(\mathbf{q},\Omega)}
{\left[\Omega-\mathrm{Re}\,\mathcal{Z}_{\mu}(\mathbf{q},\Omega)\right]^2
+\left[\mathrm{Im}\,\mathcal{Z}_{\mu}(\mathbf{q},\Omega)\right]^2}
\\[4pt]
+\, &
\frac{1}{\pi}
\frac{\mathrm{Im}\,\mathcal{Z}_{\mu}(\mathbf{q},\Omega)}
{\left[\Omega+\mathrm{Re}\,\mathcal{Z}_{\mu}(\mathbf{q},\Omega)\right]^2
+\left[\mathrm{Im}\,\mathcal{Z}_{\mu}(\mathbf{q},\Omega)\right]^2}
\Bigg],
\end{aligned}
\end{equation}
with $\mathcal{Z} _{\mu}(\mathbf{q},\Omega )=\sqrt{\omega _{\mu }^2(\mathbf{q})+\mathbf{\Pi}^{(\mathrm{B})}_{\mu \mu }(\mathbf{q},\Omega+\mathrm{i}\delta)} $. The mode-resolved spectral function is further characterized by the quasiparticle energy $\Omega_{\mu}(\mathbf{q})$ and linewidth $\Gamma_{\mu}$ within the Lorentzian approximation. These quantities can be obtained self-consistently from
\begin{equation}
\Omega_{\mu }(\mathbf{q}  )=\mathrm{Re} \mathcal{Z} _{\mu}(\mathbf{q},\Omega_ {\mu }(\mathbf{q}  )),
\end{equation}
\begin{equation}
\Gamma _{\mu }(\mathbf{q}  )=-\mathrm{Im} \mathcal{Z} _{\mu}(\mathbf{q},\Omega_ {\mu }(\mathbf{q}  )) . 
\end{equation}
In practice, for comparison with experiment, we extract the quasiparticle energies and linewidths using the one-shot approximation~\cite{bianco2018high,monacelli2021time}, where $\mathcal{Z} _{\mu}(\mathbf{q},\Omega )$ is evaluated at the auxiliary frequency $\omega_{\mu}(\mathbf{q})$, yielding  
\begin{equation}
\overset{\mathrm{(os)} }\Omega_{\mu }(\mathbf{q}  )=\mathrm{Re} \mathcal{Z} _{\mu}(\mathbf{q},\omega_ {\mu }(\mathbf{q}  )),
\end{equation}
\begin{equation}
\overset{\mathrm{(os)} }\Gamma _{\mu }(\mathbf{q}  )=-\mathrm{Im} \mathcal{Z} _{\mu}(\mathbf{q},\omega_ {\mu }(\mathbf{q}  )). 
\end{equation}

\section{LOWEST ORDER CORRECTION TO THE SSCHA FREE ENERGY}
\label{sec:cubbic}

\par As described in Section~\ref{sec:sscha}, the equilibrium SSCHA free energy is obtained by minimizing a Gaussian trial density matrix with respect to the centroid positions and the auxiliary force constants. Although the SSCHA variational free energy functional contains an infinite summation of anharmonic contributions into the optimal Gaussian reference, it is not equivalent to the complete perturbative expansion of the exact free energy. In particular, the cubic anharmonicity gives rise to an additional connected vacuum diagram, namely the bubble diagram formed by two cubic vertices connected by three one-phonon propagators, not included in the SSCHA variational free energy functional~\cite{monacelli2025analyzing}. 

\subsection{Exact residual-interaction representation}
\label{sec:ri_representation}

We work in Cartesian coordinates,
$u_a = R_a-\mathcal{R}_a$, and set
$\hbar=k_B=1$, so that $\beta=1/T$.
At the stationary SSCHA solution, the exact ionic
Hamiltonian can be decomposed as
\begin{equation}
H = \mathcal{H}_{\boldsymbol{\mathcal{R}} ,\mathbf{\Phi}}+\Delta V,
\label{eq:residual_H}
\end{equation}
where $\mathcal{H}_{\boldsymbol{\mathcal{R}}, \mathbf{\Phi}}$ is the optimized
auxiliary harmonic Hamiltonian defined in Eq.~\eqref{H}, and
\begin{equation}
\Delta V =
V_{\mathrm{BO}}
-\frac{1}{2}\sum_{ab}\Phi_{ab}u_a u_b
\label{eq:residual_V}
\end{equation}
is the residual interaction.

Since $\mathcal{H}_{\boldsymbol{\mathcal{R}}, \mathbf{\Phi}}$ and $\Delta V$
do not generally commute, the Boltzmann operator
cannot be factorized into their separate exponentials.
To construct the expansion, we introduce
$H_\lambda=\mathcal{H}_{\boldsymbol{\mathcal{R}} ,\mathbf{\Phi}}+\lambda\Delta V$,
keeping the SSCHA reference fixed.
The Duhamel formula gives
\begin{equation}
\begin{aligned}
\frac{\partial e^{-\beta H_\lambda}}{\partial\lambda}
= -\int_0^\beta d\tau\,
&e^{-(\beta-\tau)H_\lambda}\\
&\times\Delta V e^{-\tau H_\lambda}.
\end{aligned}
\label{eq:duhamel}
\end{equation}
Successive applications of this identity generate
the imaginary-time-ordered expansion.
Setting $\lambda=1$, we obtain
\begin{equation}
\begin{aligned}
e^{-\beta H}
&=e^{-\beta \mathcal{H}_{\boldsymbol{\mathcal R},\mathbf{\Phi}}}
T_\tau
\exp\left[
-\int_0^\beta d\tau\,\Delta V(\tau)
\right]
\end{aligned}
\end{equation}
and 
\begin{equation}
\begin{aligned}
\Delta V(\tau)
&=e^{\tau \mathcal{H}_{\boldsymbol{\mathcal R},\mathbf{\Phi}}}
\Delta V
e^{-\tau \mathcal{H}_{\boldsymbol{\mathcal R},\mathbf{\Phi}}},
\end{aligned}
\label{eq:boltzmann_interaction}
\end{equation}
where $T_\tau$ is the imaginary-time ordering
operator. 

Taking the trace and using the Gaussian reference
density matrix $\tilde{\rho}_{\boldsymbol{\mathcal R},\mathbf{\Phi}}$,
the exact free energy becomes~\cite{negele2018quantum}
\begin{equation}
\begin{aligned}
F=F_\mathbf{\Phi}-\frac{1}{\beta}
\ln\Bigg\langle T_\tau
\exp\Bigg[
-\int_0^\beta d\tau\,
\Delta V(\tau)
\Bigg]\Bigg\rangle_{\tilde{\rho}_{\boldsymbol{\mathcal R},\mathbf{\Phi}}},
\end{aligned}
\label{eq:exact_free_energy}
\end{equation}
where $F_\Phi$ is the auxiliary harmonic
free energy defined in Eq.~\eqref{H_F}.

Expanding the logarithm in connected cumulants yields
\begin{equation}
\begin{aligned}
    F
    &=
    F_{\mathbf{\Phi}}
    +
    \langle \Delta V\rangle_{ \tilde{\rho}_{\boldsymbol{\mathcal R},\mathbf{\Phi}}}
    \\&-
    \frac{1}{2\beta}
    \int_0^\beta d\tau
    \int_0^\beta d\tau'\,
    \left\langle
    T_\tau
    \delta\Delta V(\tau)
    \delta\Delta V(\tau')
    \right\rangle_{ \tilde{\rho}_{\boldsymbol{\mathcal R},\mathbf{\Phi}}}
    +\cdots ,
\end{aligned}
\label{eq:cumulant_free_energy}
\end{equation}
where
\(\delta\Delta V = \Delta V-\langle \Delta V\rangle_{ \tilde{\rho}_{\boldsymbol{\mathcal R},\mathbf{\Phi}}} \)
is the centered residual interaction. Here, the first two terms recover the SSCHA variational
free energy in Eq.~\eqref{F_F}. The second cumulant therefore
provides the first connected correction beyond
the Gaussian variational approximation.

\subsection{Normal ordering with respect to Gaussian state}
\label{sec:gaussian_normal_ordering}

We express the BO potential in a normal-ordered form with respect to the optimized SSCHA Gaussian state. Introducing the displacement vector $\mathbf u=\mathbf R-\boldsymbol{\mathcal R}$, with $N=3N_a$ components, we define the equal-time covariance matrix $\boldsymbol{\Psi}$ as 
\begin{equation}
    \Psi_{ab}
    =
    \langle u_a u_b\rangle_{
    \tilde\rho_{\boldsymbol{\mathcal R},\boldsymbol\Phi}}
    =
    \frac{1}{\sqrt{M_a M_b}}
   \sum_{\mu}
   \frac{\hbar (2n_{\mu}+1)}{2\omega_{\mu}}
   \, e_{\mu}^{a} e_{\mu}^{b}
    \label{eq:no_covariance}
\end{equation}
where $e_{\mu}^{a}$ is the eigenvector of mode $\mu$ along direction $a$ of the SSCHA auxiliary dynamical matrix.
For an observable $\mathds{O}(\mathbf R)$ depending only on the ionic positions, the Gaussian average reads
\begin{equation}
    \langle \mathds{O} \rangle_{
    \tilde\rho_{\boldsymbol{\mathcal R},\boldsymbol\Phi}}
    =
    \int d \mathbf{R}\,\mathds{O}(\mathbf R)
    \tilde\rho_{\boldsymbol{\mathcal R},\boldsymbol\Phi}
    (\mathbf R),
    \label{eq:no_average}
\end{equation}
where the diagonal of the ionic density matrix in the coordinate representation is
\begin{equation}
    \tilde\rho_{\boldsymbol{\mathcal R},\boldsymbol\Phi}
    (\mathbf R)
    =
    \frac{1}{\sqrt{\det(2\pi\boldsymbol{\Psi})}}
    \exp\left[
    -\frac12\mathbf u^{\mathrm T}
    \boldsymbol{\Psi}^{-1}\mathbf u
    \right].
    \label{eq:no_density}
\end{equation}

Introducing an auxiliary vector $\mathbf J=(J_1,\ldots,J_N)^{\mathrm T}$, the multivariate Gaussian generating function is
\begin{equation}
    \mathcal Z(\mathbf J)
    \equiv
    \left\langle e^{\mathbf J^{\mathrm T}\mathbf u}
    \right\rangle_{
    \tilde\rho_{\boldsymbol{\mathcal R},\boldsymbol\Phi}}
    =
    \exp\left(
    \frac12\mathbf J^{\mathrm T}
    \boldsymbol{\Psi}\mathbf J
    \right).
    \label{eq:no_Z}
\end{equation}
The normal-ordering generating function is defined by
\begin{equation}
    \mathcal N(\mathbf J,\mathbf u)
    \equiv
    \frac{e^{\mathbf J^{\mathrm T}\mathbf u}}{\mathcal Z(\mathbf J)}
    =
    \exp\left[
    \mathbf J^{\mathrm T}\mathbf u
    -\frac12\mathbf J^{\mathrm T}
    \boldsymbol{\Psi}\mathbf J
    \right].
    \label{eq:no_N}
\end{equation}
The normal-ordered displacement products are
obtained from
\begin{equation}
    :u_{a_1}\cdots u_{a_n}:_{
    \tilde\rho_{\boldsymbol{\mathcal R},\boldsymbol\Phi}}
    =
    \left.
    \frac{\partial^n\mathcal N(\mathbf J,\mathbf u)}
    {\partial J_{a_1}\cdots\partial J_{a_n}}
    \right|_{\mathbf J=\mathbf 0}.
    \label{eq:no_definition}
\end{equation}
For example,
\begin{equation}
    :u_a:_{\tilde\rho_{\boldsymbol{\mathcal R},\boldsymbol\Phi}}
    =u_a,
\end{equation}
\begin{equation}
    :u_a u_b:_{\tilde\rho_{\boldsymbol{\mathcal R},\boldsymbol\Phi}}
    =u_a u_b-\Psi_{ab},
\end{equation}
\begin{align}
    &:u_a u_b u_c:_{
    \tilde\rho_{\boldsymbol{\mathcal R},\boldsymbol\Phi}}
    \nonumber\\
    &=
    u_a u_b u_c-\Psi_{ab}u_c
    -\Psi_{ac}u_b-\Psi_{bc}u_a.
\end{align}
At fourth order,
\begin{align}
    &:u_a u_b u_c u_d:_{
    \tilde\rho_{\boldsymbol{\mathcal R},\boldsymbol\Phi}}
    \nonumber\\
    &=
    u_a u_b u_c u_d
    -\Psi_{ab}u_cu_d
    -\Psi_{ac}u_bu_d
    \nonumber\\
    &\quad
    -\Psi_{ad}u_bu_c
    -\Psi_{bc}u_au_d
    \nonumber\\
    &\quad
    -\Psi_{bd}u_au_c
    -\Psi_{cd}u_au_b
    \nonumber\\
    &\quad
    +\Psi_{ab}\Psi_{cd}
    +\Psi_{ac}\Psi_{bd}
    +\Psi_{ad}\Psi_{bc}.
    \label{eq:no_fourth}
\end{align}
All normal-ordered products of positive rank have vanishing Gaussian averages,
\begin{align}
    &\left\langle
    :u_{a_1}\cdots u_{a_n}:_{
    \tilde\rho_{\boldsymbol{\mathcal R},\boldsymbol\Phi}}
    \right\rangle_{
    \tilde\rho_{\boldsymbol{\mathcal R},\boldsymbol\Phi}}
    =0, n\geq1.
    \label{eq:no_zero}
\end{align}
Furthermore, normal-ordered products belonging to different Gaussian polynomial ranks are
mutually orthogonal,
\begin{equation}
    \Big\langle
    :u_{a_1}\cdots u_{a_n}:_{
    \tilde\rho_{\boldsymbol{\mathcal R},\boldsymbol\Phi}}
    \times
    :u_{b_1}\cdots u_{b_m}:_{
    \tilde\rho_{\boldsymbol{\mathcal R},\boldsymbol\Phi}}
    \Big\rangle_{
    \tilde\rho_{\boldsymbol{\mathcal R},\boldsymbol\Phi}}
    =0, n\neq m.
    \label{eq:no_orthogonality}
\end{equation}
These relations are derived in
Appendix~\ref{app:normal_ordering}.

\subsection{Normal-ordered BO potential and residual vertices}
\label{sec:normal_ordered_BO}

The orthogonality of different Gaussian normal-ordered
ranks allows the BO potential to be expanded in the
corresponding Hermite polynomial basis. The expansion
coefficients are obtained by projecting the BO potential
onto the normal-ordered displacement products,
\begin{align}
    \Phi^{(n)}_{a_1\cdots a_n}
    ={}&
    \sum_{b_1\cdots b_n}
    \prod_{i=1}^{n}(\Psi^{-1})_{a_i b_i}
    \nonumber\\
    &\times
    \Big\langle V_{\mathrm{BO}}
    :u_{b_1}\cdots u_{b_n}:_{
    \tilde\rho_{\boldsymbol{\mathcal R},\boldsymbol\Phi}}
    \Big\rangle_{
    \tilde\rho_{\boldsymbol{\mathcal R},\boldsymbol\Phi}}
    \nonumber\\
    ={}&
    \left\langle
    \frac{\partial^n V_{\mathrm{BO}}}
    {\partial u_{a_1}\cdots\partial u_{a_n}}
    \right\rangle_{
    \tilde\rho_{\boldsymbol{\mathcal R},\boldsymbol\Phi}}.
    \label{eq:BO_averaged_vertices}
\end{align}
The second equality follows from Gaussian integration
by parts, as derived in Appendix~\ref{app:BO_projection}.

The full BO potential can therefore be formally
reorganized as
\begin{align}
    V_{\mathrm{BO}}
    ={}&
    \langle V_{\mathrm{BO}}\rangle_{
    \tilde\rho_{\boldsymbol{\mathcal R},\boldsymbol\Phi}}
    \nonumber\\
    &+
    \sum_{n=1}^{\infty}\frac{1}{n!}
    \sum_{a_1\cdots a_n}
    \Phi^{(n)}_{a_1\cdots a_n}
    \nonumber\\
    &\qquad\times
    :u_{a_1}\cdots u_{a_n}:_{
    \tilde\rho_{\boldsymbol{\mathcal R},\boldsymbol\Phi}}.
    \label{eq:BO_normal_expansion}
\end{align}
Using the normal-ordered expansion, the fluctuation of the residual interaction \(\delta\Delta V\) reads
\begin{equation}
\begin{aligned}
    \delta\Delta V
    =
    &
    \sum_a
    \Phi^{(1)}_a
    :u_a:_{ \tilde{\rho}_{\boldsymbol{\mathcal R},\mathbf{\Phi}}}
    \\
    &
    +
    \frac12
    \sum_{ab}
    \left[
    \Phi^{(2)}_{ab}
    -
    \Phi_{ab}
    \right]
    :u_a u_b:_{ \tilde{\rho}_{\boldsymbol{\mathcal R},\mathbf{\Phi}}}
    \\
    &
    +
    \frac{1}{3!}
    \sum_{abc}
    \Phi^{(3)}_{abc}
    :u_a u_b u_c:_{ \tilde{\rho}_{\boldsymbol{\mathcal R},\mathbf{\Phi}}}
    +\cdots .
\end{aligned}
\end{equation}

\begin{figure}[t]
    \centering
    \includegraphics[width=\linewidth]{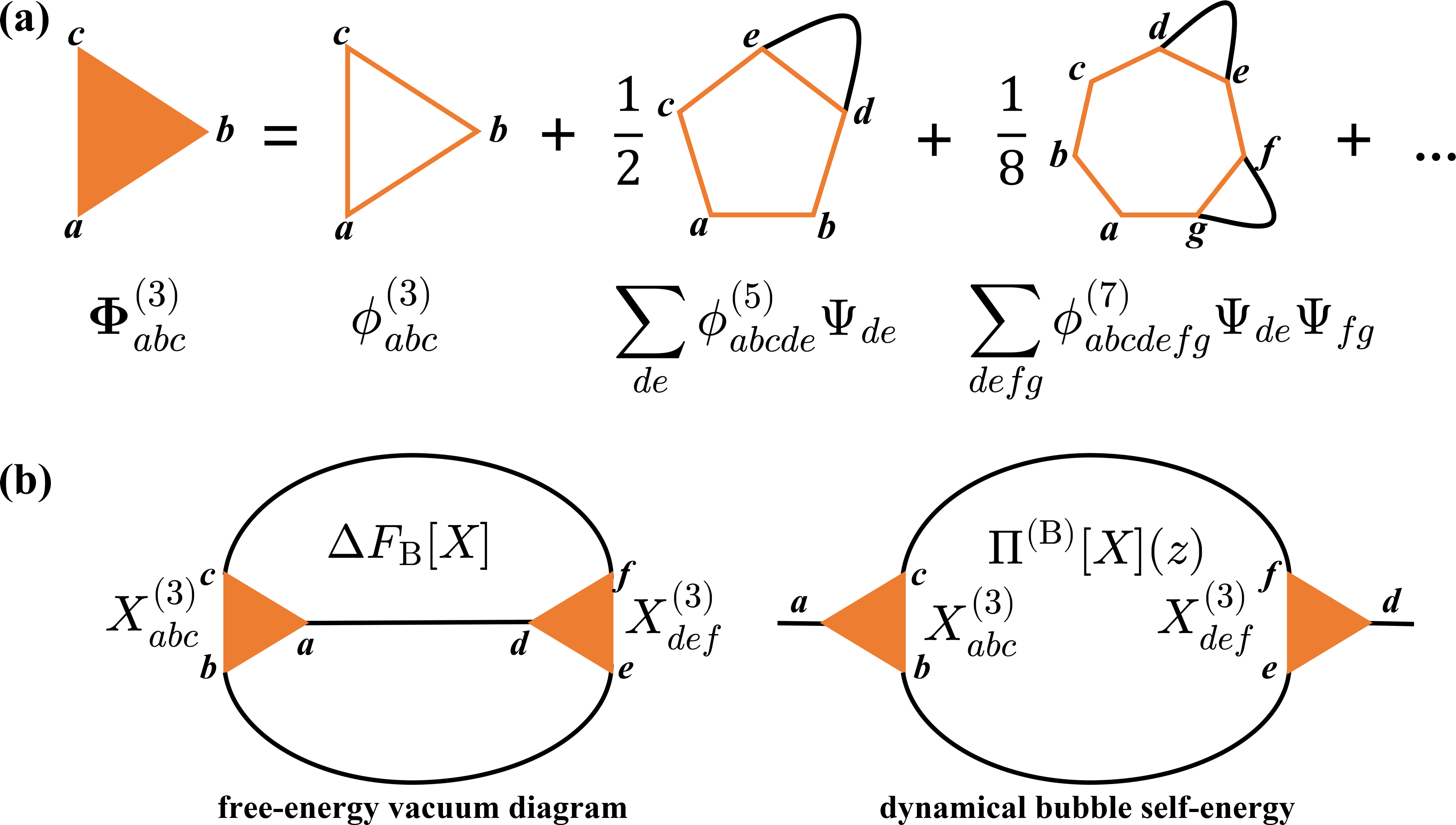}
    \caption{\label{Fig1} Gaussian-averaged cubic vertex and its diagrammatic contributions. (a) The Gaussian-averaged cubic vertex \(\Phi_{abc}^{(3)}\) can be formally expressed in terms of the bare BO cubic derivative \(\phi_{abc}^{(3)}\) and Gaussian contractions of higher-order odd derivatives with the equal-time covariance \(\Psi\). (b) Two cubic vertices connected by three auxiliary propagators \(G^S\) give rise to the free-energy vacuum correction \(\Delta F_{\mathrm B}\), while opening one propagator line yields the corresponding dynamical bubble self-energy \(\Pi^{(\mathrm B)}(z)\). The cubic vertices labeled \(X^{(3)}\) in panel (b) represent either the Gaussian-averaged vertex \(\Phi^{(3)}\) or the bare vertex \(\phi^{(3)}\). 
    }
\end{figure} 

At the fully stationary solution, i.e. at the minimum of the SSCHA minimization,
\begin{equation}
    \Phi^{(1)}_a
    =
    \left\langle
    \frac{\partial V_{\rm BO}}{\partial u_a}
    \right\rangle_{ \tilde{\rho}_{\boldsymbol{\mathcal R},\mathbf{\Phi}}}
    =
    0,
\end{equation}
and
\begin{equation}
    \Phi^{(2)}_{ab}
    =
    \left\langle
    \frac{\partial^2 V_{\rm BO}}
    {\partial u_a\partial u_b}
    \right\rangle_{ \tilde{\rho}_{\boldsymbol{\mathcal R},\mathbf{\Phi}}}
    =
    \Phi_{ab}.
    \label{eq:second_order}
\end{equation}
Therefore, the linear and quadratic residual vertices vanish in the normal-ordered representation. The lowest-rank nonvanishing residual term is cubic,
\begin{equation}
    \delta\Delta V
    =
    \frac{1}{3!}
    \sum_{abc}
    \Phi^{(3)}_{abc}
    :u_au_bu_c:_{ \tilde{\rho}_{\boldsymbol{\mathcal R},\mathbf{\Phi}}}
    +\cdots,
    \label{eq:residual_cubic}
\end{equation}
where 
\begin{align}
    \Phi^{(3)}_{abc}
    &=
    \left\langle
    \frac{\partial^3 V_{\rm BO}}
    {\partial u_{a} \partial u_{b} \partial u_{c}}
    \right\rangle_{ \tilde{\rho}_{\boldsymbol{\mathcal R},\mathbf{\Phi}}}
    \nonumber\\
    &=
    \sum_{m=0}^{\infty}
    \frac{1}{2^m m!}
    \phi^{(3+2m)}_{abc a_1\cdots a_{2m}}
    \prod_{j=1}^{m}
    \Psi_{a_{2j-1}a_{2j}}
    \nonumber\\
    &=
    \phi^{(3)}_{abc}
    +
    \frac{1}{2}  \sum_{de}
    \phi^{(5)}_{abcde}\Psi_{de}
    +
    \frac{1}{8}  \sum_{defg}
    \phi^{(7)}_{abcdefg}
    \Psi_{de}\Psi_{fg}
    +\cdots ,
    \label{eq:Phi_3}
\end{align}
In Eq.~\eqref{eq:Phi_3}, we explicitly show the dependence of the third-order average vertex on high-order bare vertices calculated at the equilibrium SSCHA position,
\begin{equation}
    \phi^{(n)}_{a_1\cdots a_n}
    =
    \left[
    \frac{\partial^n V_{\rm BO}}
    {\partial u_{a_1}\cdots\partial u_{a_n}}
    \right]_{\boldsymbol{\mathcal{R}}_{\mathrm{eq} }}.
\end{equation}
Fig.~\ref{Fig1}(a) shows a diagrammatic representation of the third-order average vertex making explicit its dependence on high-order bare anharmonic terms.

\subsection{Cubic vacuum-bubble correction to the SSCHA
free energy}
\label{sec:cubic_bubble}

Retaining the normal-ordered cubic residual interaction
in the second cumulant gives the cubic vacuum-bubble
correction to the SSCHA free energy,
\begin{equation}
\begin{aligned}
\Delta F_{\mathrm B}
={}&-\frac{1}{2\beta}
\int_0^\beta d\tau
\int_0^\beta d\tau'
\\
&\times
\left\langle
T_\tau
\delta\Delta V_3(\tau)
\delta\Delta V_3(\tau')
\right\rangle_{
\tilde{\rho}_{\boldsymbol{\mathcal R},\mathbf{\Phi}}},
\end{aligned}
\label{eq:cubic_vacuum_initial}
\end{equation}
where $\delta\Delta V_3$ denotes the cubic term
of the centered residual interaction defined in
Eq.~\eqref{eq:residual_cubic}. Eq.~\eqref{eq:cubic_vacuum_initial} is represented diagrammatically by the left-hand vacuum diagram in Fig.~\ref{Fig1}(b).

Since the cubic vertices are normal ordered with
respect to the optimized Gaussian reference,
all contractions within the same vertex vanish.
Wick's theorem therefore leaves only contractions
connecting the two vertices.
Using the permutation symmetry of the cubic
vertices and imaginary-time translational invariance,
we transform the resulting expression to the
normal-mode basis of the auxiliary Hamiltonian.

The mass-weighted cubic vertex in this basis is
defined as
\begin{equation}
D_{\mu\nu\lambda}^{(3)}
=
\sum_{abc}
\frac{\Phi_{abc}^{(3)}}
{\sqrt{M_aM_bM_c}}
e_a^\mu e_b^\nu e_c^\lambda.
\label{eq:cubic_vertex_mode}
\end{equation}
The free-energy correction then becomes
\begin{equation}
\Delta F_{\mathrm B}
=
-\frac{1}{12}
\sum_{\mu\nu\lambda}
\left|D_{\mu\nu\lambda}^{(3)}\right|^2
\int_0^\beta d\tau\,
G_\mu^S(\tau)
G_\nu^S(\tau)
G_\lambda^S(\tau).
\label{eq:cubic_vacuum_integral}
\end{equation}
Here $G_\mu^S(\tau)$ is the auxiliary
imaginary-time propagator, given for
$0<\tau<\beta$ by
\begin{equation}
\begin{aligned}
G_\mu^S(\tau)
&=
\frac{1}{2\omega_\mu}
\left[
(1+n_\mu)e^{-\omega_\mu\tau}
+n_\mu e^{\omega_\mu\tau}
\right],
\\
n_\mu&=\frac{1}{e^{\beta\omega_\mu}-1}.
\end{aligned}
\label{eq:cubic_auxiliary_propagator}
\end{equation}

Performing the imaginary-time integral in
Eq.~\eqref{eq:cubic_vacuum_integral} and using
the permutation symmetry of the cubic vertices
yields
\begin{equation}
\begin{aligned}
\Delta F_{\mathrm B}
={}&-\frac{1}{48}
\sum_{\mu\nu\lambda}
\frac{\left|D_{\mu\nu\lambda}^{(3)}\right|^2}
{\omega_\mu\omega_\nu\omega_\lambda}
\\
&\times
\Bigg[
\frac{
(1+n_\mu)(1+n_\nu)(1+n_\lambda)
-n_\mu n_\nu n_\lambda
}{
\omega_\mu+\omega_\nu+\omega_\lambda
}
\\
&\qquad+
3\frac{
n_\mu(1+n_\nu+n_\lambda)-n_\nu n_\lambda
}{
-\omega_\mu+\omega_\nu+\omega_\lambda
}
\Bigg].
\end{aligned}
\label{eq:cubic_vacuum_final}
\end{equation}
The sum runs over all ordered auxiliary
normal-mode triplets.
The factor of three in the second term follows
from the permutation symmetry of the cubic
vertex under the complete mode summation.
The Wick contractions and the explicit
imaginary-time integration are detailed in
Appendix~\ref{app:cubic_bubble}.

Eq.~\eqref{eq:cubic_vacuum_final}
represents the lowest-rank contribution to the
second cumulant around the optimized SSCHA
Gaussian reference. Unlike the bare cubic
vertex, the Gaussian-averaged vertex
$\Phi^{(3)}$ includes the effects of higher-order
odd derivatives of the full BO potential,
as illustrated in Fig.~\ref{Fig1}(a).
For a BO potential truncated at fourth order,
the averaged cubic vertex reduces to the
bare cubic derivative,
$\Phi^{(3)}=\phi^{(3)}$.
The apparent singularities associated with
vanishing frequency-difference denominators
are removable and have finite limiting values.

\section{ANHARMONIC MODE-RESOLVED GR\"UNEISEN PARAMETER}
\label{app:sscha_gruneisen}
We derive a mode-resolved thermal-expansion formula from the variational SSCHA free energy, in which the key step is to express the ionic entropy in terms of the optimized auxiliary phonon modes and relate its volume derivative to thermal expansion. In this section, $V$ is the primitive-cell volume, varied by isotropically scaling the cubic cell vectors $\{\vec{a}_i\}$. 

Using Eq.~\eqref{F_F}, we can write the SSCHA variational functional as
\begin{equation}
\begin{aligned}
F(\boldsymbol{\mathcal R},\boldsymbol\Phi;V,T)
={}& F_{\boldsymbol\Phi}(T)
+\left\langle V_{\mathrm{BO}}-V_{\boldsymbol{\mathcal R},\boldsymbol\Phi}
\right\rangle_{\tilde\rho_{\boldsymbol{\mathcal R},\boldsymbol\Phi}},
\\
V_{\boldsymbol{\mathcal R},\boldsymbol\Phi}
={}&\frac12\sum_{ab}\Phi_{ab}u_a u_b,
\qquad u_a=R_a-\mathcal R_a.
\end{aligned}
\label{eq:app_gr_var}
\end{equation}
The minimum free energy is 
\begin{equation}
F_{\tilde\rho_{\boldsymbol{\mathcal R},\boldsymbol\Phi}}(V,T)
=F(\boldsymbol{\mathcal R}_{\mathrm{eq}},
\boldsymbol\Phi_{\mathrm{eq}};V,T),
\label{eq:app_gr_min}
\end{equation}
with the stationarity conditions
\begin{equation}
\left.\frac{\partial F}{\partial\mathcal R_a}\right|_{\mathrm{eq}}=0,
\qquad
\left.\frac{\partial F}{\partial\Phi_{ab}}\right|_{\mathrm{eq}}=0.
\label{eq:app_gr_stationarity}
\end{equation}
We first show that the entropy of the optimized SSCHA state has the same functional form as that of its auxiliary harmonic Hamiltonian. Differentiating Eq.~\eqref{eq:app_gr_min} with respect to temperature
at fixed volume gives
\begin{equation}
\begin{aligned}
\left(\frac{\partial F_{\tilde\rho_{\boldsymbol{\mathcal R},
\boldsymbol\Phi}}}{\partial T}\right)_V
={}&\left.\frac{\partial F}{\partial T}\right|_{\boldsymbol{\mathcal R},
\boldsymbol\Phi,V}
+\sum_a\frac{\partial F}{\partial\mathcal R_a}
\left(\frac{\partial\mathcal R_{\mathrm{eq},a}}{\partial T}\right)_V
\\
&+\sum_{ab}\frac{\partial F}{\partial\Phi_{ab}}
\left(\frac{\partial\Phi_{\mathrm{eq},ab}}{\partial T}\right)_V
\\
={}&\left.\frac{\partial F}{\partial T}\right|_{\boldsymbol{\mathcal R},
\boldsymbol\Phi,V,\mathrm{eq}},
\end{aligned}
\label{eq:app_gr_envelope}
\end{equation}
Here, Eq.~\eqref{eq:app_gr_stationarity} makes the two implicit derivatives vanish. The remaining explicit temperature derivative continues to act on the Gaussian averages from Eq.~\eqref{eq:app_gr_var}.

From Eq.~\eqref{eq:no_covariance}, the equal-time covariance matrix is given by $\Psi_{ab}=\langle u_a u_b\rangle_{\tilde\rho_{\boldsymbol{\mathcal R}, \boldsymbol\Phi}}$. Using the multivariate Gaussian integration-by-parts identity (Appendix~\ref{app:BO_projection}) with fixed $\boldsymbol{\mathcal R}$,
$\boldsymbol\Phi$, and $V$,
\begin{equation}
\begin{aligned}
\left.\frac{\partial}{\partial T}
\langle V_{\mathrm{BO}}\rangle_{\tilde\rho_{\boldsymbol{\mathcal R},
\boldsymbol\Phi}}\right|_{\boldsymbol{\mathcal R},\boldsymbol\Phi,V}
={}&\frac12\sum_{ab}
\left\langle\frac{\partial^2 V_{\mathrm{BO}}}
{\partial u_a\partial u_b}\right\rangle_{\tilde\rho_{\boldsymbol{\mathcal R},
\boldsymbol\Phi}}
\\
&\times\left.\frac{\partial\Psi_{ab}}{\partial T}
\right|_{\boldsymbol{\mathcal R},\boldsymbol\Phi,V}.
\end{aligned}
\label{eq:app_gr_gaussian_derivative}
\end{equation}
For the auxiliary potential,
$\langle V_{\boldsymbol{\mathcal R},\boldsymbol\Phi}\rangle_{
\tilde\rho_{\boldsymbol{\mathcal R},\boldsymbol\Phi}}
=\frac12\sum_{ab}\Phi_{ab}\Psi_{ab}$, we have
\begin{equation}
\begin{aligned}
\left.\frac{\partial F}{\partial T}
\right|_{\boldsymbol{\mathcal R},\boldsymbol\Phi,V}
={}&\left.\frac{\partial F_{\boldsymbol\Phi}}{\partial T}
\right|_{\boldsymbol\Phi}
\\
&+\frac12\sum_{ab}
\left(\Phi_{ab}^{(2)}-\Phi_{ab}\right)
\left.\frac{\partial\Psi_{ab}}{\partial T}
\right|_{\boldsymbol{\mathcal R},\boldsymbol\Phi,V}.
\end{aligned}
\label{eq:app_gr_temperature_derivative}
\end{equation}
Here, $\Phi_{ab}^{(2)}$ represents the Gaussian-averaged BO second derivative from Eq.~\eqref{eq:second_order}. At the stationary SSCHA solution, $\Phi_{ab}^{(2)}=\Phi_{ab}$, the second term in Eq.~\eqref{eq:app_gr_temperature_derivative} vanishes. Substituting the auxiliary harmonic density matrix relation $-(\partial F_{\boldsymbol\Phi}/\partial T)_{\boldsymbol\Phi}
=S_{\mathrm{ion}}[\tilde\rho_{\boldsymbol{\mathcal R},\boldsymbol\Phi}]$ into this result directly yields
\begin{equation}
-\left(\frac{\partial F_{\tilde\rho_{\boldsymbol{\mathcal R},
\boldsymbol\Phi}}}{\partial T}\right)_V
=S_{\mathrm{ion}}[\tilde\rho_{\boldsymbol{\mathcal R}_{\mathrm{eq}},
\boldsymbol\Phi_{\mathrm{eq}}}].
\label{eq:app_gr_entropy_identity}
\end{equation}

Equation~(\ref{eq:app_gr_entropy_identity}) shows that the entropy obtained from the temperature derivative of the minimized SSCHA free energy equals the von Neumann entropy of the optimized Gaussian density matrix.  Because the optimized SSCHA Gaussian state is defined by the auxiliary harmonic Hamiltonian $\mathcal{H}_{\boldsymbol{\mathcal{R}} ,\mathbf{\Phi}}$, we can express the entropy in terms of the auxiliary phonon frequencies. Denoting these frequencies by $\omega_\mu(V,T)$,
the Bose--Einstein occupation number is $n_\mu=(e^{\beta\hbar\omega_\mu}-1)^{-1}$,
where $\beta=(k_BT)^{-1}$. Eq.~\eqref{eq:app_gr_entropy_identity} then gives
\begin{equation}
\begin{aligned}
S_{\mathrm{ion}}(V,T)
=
k_B\sum_\mu
\big[
&(n_\mu+1)\ln(n_\mu+1)
\\
&-n_\mu\ln n_\mu
\big].
\end{aligned}
\label{eq:app_gr_entropy_modes}
\end{equation}
To relate the entropy to thermal expansion, we introduce the SSCHA pressure
$P_{\rm S}=-(\partial F_{\tilde\rho_{\boldsymbol{\mathcal R}, \boldsymbol\Phi}}/\partial V)_T$ and the isothermal bulk modulus $B=-V(\partial P_{\rm S}/\partial V)_T$. For cubic $\mathrm{ScF}_3$, the linear thermal-expansion coefficient is
\begin{equation}
\alpha(T)
=
\frac{1}{a_0(T)}\frac{da_0(T)}{dT}
=
\frac{1}{3V_0(T)}\frac{dV_0(T)}{dT},
\label{eq:app_gr_alpha_definition}
\end{equation}
where $V_0(T)$ is the zero-pressure equilibrium volume. Differentiating the equilibrium condition
$P_S(V_0(T),T)=0$ and using the definition of
$B$, we obtain
\begin{equation}
\begin{aligned}
3B\alpha(T)
=
\left(\frac{\partial P_S}{\partial T}\right)_V
&=
-\frac{\partial^2
F_{\tilde\rho_{\boldsymbol{\mathcal R},
\boldsymbol\Phi}}}{\partial V\partial T}
\\
&=
\left(\frac{\partial S_{\mathrm{ion}}}{\partial V}
\right)_T.
\end{aligned}
\label{eq:app_gr_maxwell}
\end{equation}
Here the last equality follows from the Maxwell relation. From Eq.~\eqref{eq:app_gr_entropy_modes}, the volume derivative of the entropy can be written as
\begin{equation}
\begin{aligned}
\left(\frac{\partial S_{\mathrm{ion}}}{\partial V}
\right)_T
&=
\sum_\mu
\left(\frac{\partial s_\mu}{\partial\omega_\mu}
\right)_T
\left(\frac{\partial\omega_\mu}{\partial V}
\right)_T
\\
&=
-\sum_\mu\frac{c_\mu}{\omega_\mu}
\left(\frac{\partial\omega_\mu}{\partial V}
\right)_T,
\end{aligned}
\label{eq:app_gr_entropy_volume}
\end{equation}
where $s_\mu$ is the single-mode entropy and
\begin{equation}
c_\mu(V,T)
=
k_B(\beta\hbar\omega_\mu)^2
n_\mu(n_\mu+1)
\label{eq:app_gr_mode_c}
\end{equation}
is the auxiliary mode heat-capacity weight.

We therefore define the mode-resolved SSCHA Gr\"uneisen parameter as
\begin{equation}
\gamma_\mu(V,T)
=
-\frac{V}{\omega_\mu(V,T)}
\left(\frac{\partial\omega_\mu(V,T)}
{\partial V}\right)_T,
\label{eq:app_gr_gamma_definition}
\end{equation}
and obtain the linear thermal-expansion coefficient as
\begin{equation}
\alpha(T)
=
\frac{1}{3B V_0}
\sum_\mu
c_\mu(V_0,T)\gamma_\mu(V_0,T).
\label{eq:app_gr_alpha}
\end{equation}

To identify the vibrational contributions to
thermal expansion, we introduce
\begin{align}
\alpha_\mu(T)
&=
\frac{c_\mu(V_0,T)\gamma_\mu(V_0,T)}
{3BV_0},
\label{eq:app_gr_mode_alpha}
\\
\alpha_\pm(T)
&=
\sum_{\pm\gamma_\mu(V_0,T)>0}
\alpha_\mu(T),
\label{eq:app_gr_positive_negative}
\\
\alpha_{<\omega_c}(T)
&=
\sum_\mu
\Theta\!\left(
\omega_c-\omega_\mu(V_0,T)
\right)\alpha_\mu(T),
\label{eq:app_gr_cumulative}
\end{align}
where $\alpha(T)=\sum_\mu\alpha_\mu(T)=\alpha_+(T)+\alpha_-(T)$, and $\Theta$ is the step function. The quantities $\alpha_\pm(T)$ represent the total positive and negative contributions, $\Delta\alpha_\pm(\omega_c,T)$ their frequency-resolved contributions, and
$\alpha_{<\omega_c}(T)$ the cumulative contribution.

\section{COMPUTATIONAL DETAILS}
\label{sec:methods}

\subsection{First-principles calculations}
\par Density-functional theory (DFT) and density-functional perturbation theory (DFPT)~\cite{baroni2001phonons} calculations were performed using the \textsc{Quantum ESPRESSO} package~\cite{giannozzi2009quantum, giannozzi2017advanced}. The exchange-correlation energy was treated within the local-density approximation (LDA)~\cite{perdew1992accurate}. The electron-ion interaction was described using optimized norm-conserving Vanderbilt (ONCV) pseudopotentials~\cite{hamann2013optimized} generated consistently within LDA and obtained from the PseudoDojo library~\cite{van2018pseudodojo}. The Kohn-Sham wave functions and charge density were expanded using kinetic-energy cutoffs of 100 and 400~Ry, respectively, for the plane-waves basis. Brillouin-zone (BZ) integrations were performed using a $12\times12\times12$ Monkhorst--Pack $k$-point mesh with fixed electronic occupations. The self-consistent calculations employed Davidson diagonalization with an energy-convergence threshold of $10^{-10}$~Ry.

\subsection{SSCHA calculations}
\par Stochastic self-consistent harmonic approximation calculations were performed using \(4\times4\times4\) and \(3\times3\times3\) supercells for the \(Pm\bar{3}m\) and \(R\bar{3}c\) phases, respectively. The cubic $Pm\bar{3}m$ phase was treated entirely at the DFT level. The energies, forces, and stresses of all stochastic configurations, as well as the free-energy Hessian, were evaluated directly from DFT. The \(Pm\bar{3}m\) centroid minimizations used 200 configurations per population, while convergence of the free-energy Hessian required 1000 DFT configurations. The \(R\bar{3}c\) calculations were used only to follow the symmetry-broken structural relaxation and were accelerated using a DFT-trained Deep Potential model~\cite{wang2018deepmd,zeng2023deepmd} (see Appendix~\ref{app:dp} for Deep Potential training and validation).

\par The optimized SSCHA auxiliary force constants were Fourier interpolated onto uniform reciprocal-space meshes for the evaluation of the free energy, Hessian, and dynamical quantities. The vibrational mode sum entering the SSCHA variational free energy was evaluated on a $12\times12\times12$ mesh. Its convergence was assessed at 50~K from the corresponding free-energy equations of state (see Appendix~\ref{app:qmesh_convergence} for convergence of the free-energy equation of state). A denser $24\times24\times24$ mesh was used to evaluate the Brillouin-zone sums entering the SSCHA free-energy Hessian. The dynamical bubble self-energy, anharmonic spectral functions, and quasiparticle energies were evaluated using a $12\times12\times12$ mesh. The retarded Green function was calculated at the complex frequency $\omega+i\delta$, with $\delta=10$~cm$^{-1}$. The Dirac delta peaks of the harmonic reference spectra were represented by Gaussian functions with the same numerical broadening of 10~cm$^{-1}$. For the mode-resolved analysis of the $M_3^+$ and $R_4^+$ soft modes, we adopted the no-mode-mixing approximation and retained only the diagonal components of the dynamical bubble self-energy (see Appendix~\ref{app:bubble} for a comparison of the full and bubble-approximated Hessians) in the SSCHA phonon-mode basis. The quasiparticle energies were extracted within the one-shot approximation by evaluating the self-energy at the corresponding auxiliary frequencies. The cubic vacuum-bubble free-energy correction $\Delta F_{\rm B}$ was evaluated on the same $12\times12\times12$
mesh at the sampled volumes. At each temperature, its volume dependence was represented by a centered quadratic interpolation (see Appendix~\ref{app:bubble_volume_fit} for the interpolation procedure and its validation).

\subsection{SCP calculations}
The second- and higher-order interatomic force constants (IFCs) required for the SCP calculations were determined using a real-space $4\times4\times4$ supercell. The  second-order IFCs were extracted from the resulting displacement-force dataset by least-squares fitting, as implemented in \textsc{Alamode}~\cite{tadano2014anharmonic, tadano2015self}. The higher-order IFCs were determined independently by fitting a Taylor expansion potential (TEP). The TEP was truncated at sixth order in atomic displacements while only the second- and fourth-order IFCs were used in the SCP calculations. At each sampled volume, 1000 configurations were generated by Monte Carlo sampling, and the corresponding atomic forces were evaluated from first-principles calculations. The
anharmonic IFCs were extracted making use of the compressive sensing lattice dynamics method~\cite{zhou2014lattice, zhou2019compressive, zhou2019compressive2} employing the least absolute shrinkage and selection operator (LASSO), with the regularization parameter optimized by cross-validation. During the LASSO regression, the second-order IFCs were fixed, and only the higher-order terms were optimized. The SCP equations were subsequently solved using a $12\times12\times12$ mesh. The same mesh was employed for all internal summations, the interpolation of the effective dynamical matrices, the calculation of the renormalized phonon frequencies, and the evaluation of the SCP variational free energy.

\begin{figure}[t]
    \centering
    \includegraphics[width=\linewidth]{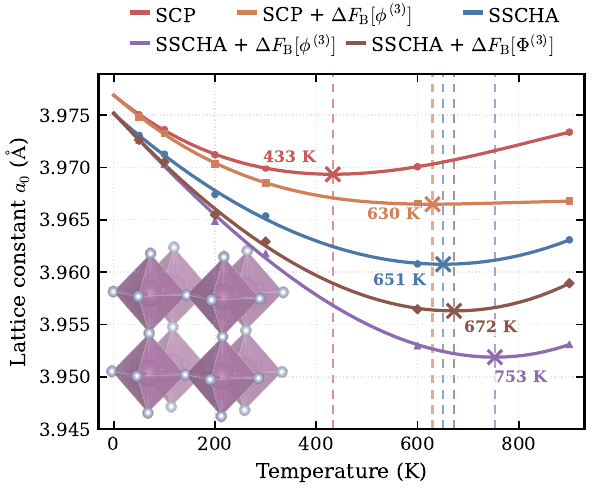}
    \caption{\label{Fig2} Equilibrium lattice constant $a_0(T)$ from SCP, SSCHA, and their corresponding cubic vacuum-bubble corrections. Filled symbols represent the calculated data points used to fit the solid curves. Crosses mark the points on the fitted curves where the thermal-expansion coefficient vanishes, $\alpha(T^*)=0$, while the corresponding colored vertical dashed lines indicate the crossover temperatures $T^*$, to be compared with the experimental value of approximately 1100 K~\cite{greve2010pronounced}. The inset shows cubic ScF$_3$ as corner-sharing ScF$_6$ octahedra (purple), with Sc atoms at their centers and F atoms at their vertices.
}
\end{figure} 

\begin{figure}[b]
    \centering
    \includegraphics[width=\linewidth]{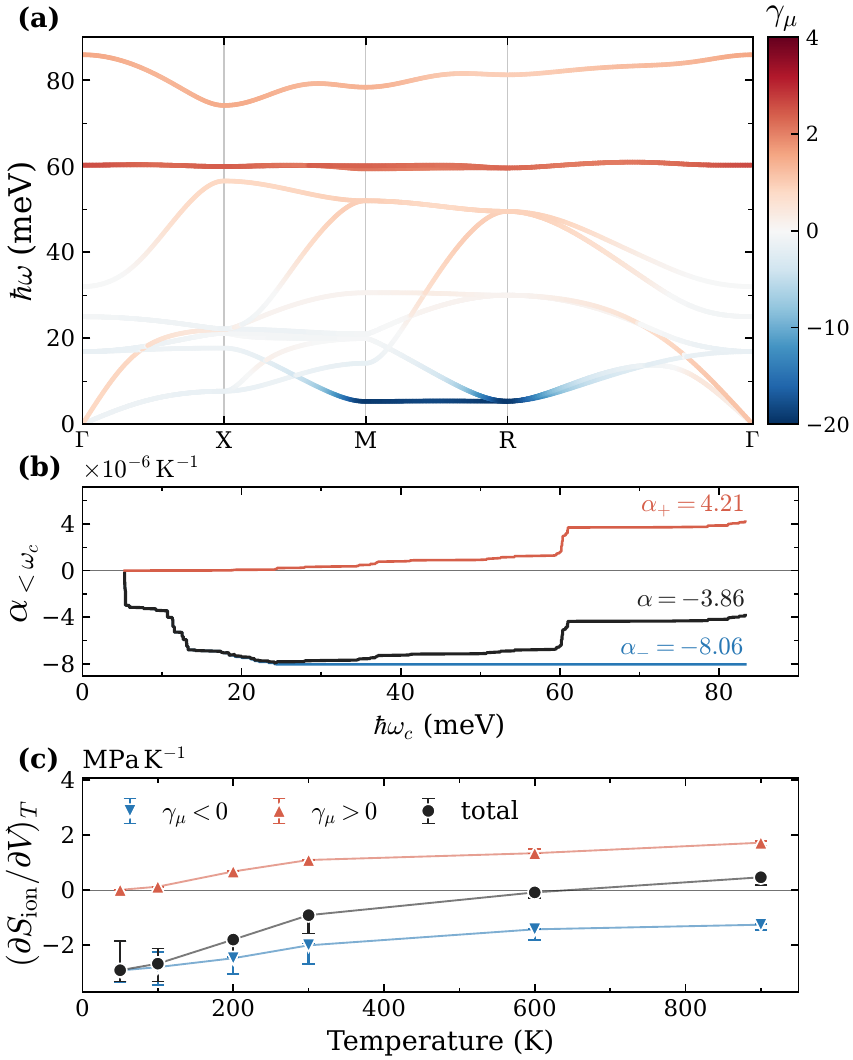}
    \caption{\label{fig:gruneisen} Mode-resolved Gr\"uneisen analysis of cubic ScF$_3$. (a) SSCHA auxiliary phonon dispersion at 300 K, colored by the mode Gr\"uneisen parameter $\gamma_\mu$. (b)    Cumulative positive ($\alpha_+$), negative ($\alpha_-$), and total ($\alpha$) contributions to the linear thermal expansion coefficient as a function of the phonon energy cutoff $\hbar\omega_c$. (c) Temperature dependence of the positive, negative, and total contributions to the entropy volume derivative $(\partial S_{\mathrm{ion}}/\partial V)_T$, evaluated at the equilibrium volumes $V_0(T)$ obtained from the SSCHA free-energy equation of state. 
}
\end{figure} 

\section{RESULTS AND DISCUSSION IN SCANDIUM TRIFLUORIDE}
\label{sec:results}
\subsection{Thermal-expansion crossover and mode-resolved contributions}
\label{sec:nte}
\par Fig.~\ref{Fig2} shows the equilibrium lattice constant \(a_0(T)=\arg\underset{a}{\min}F(a,T)\) at zero external pressure. Because it is determined by the minimum of the free-energy, $a_0(T)$ provides a sensitive probe of both the
non-perturbative SSCHA result as well as  the cubic vacuum-bubble
correction to the free energy. We compare five levels of approximation. In the SCP  calculation an auxiliary Gaussian state is optimized for the quartic-truncated potential \(V_{\rm BO\leq 4}\). Adding the cubic vacuum diagram evaluated with the obtained result gives SCP\(+\Delta F_{\rm B}[\phi^{(3)}]\). As noted above, the truncation of the BO potential means that the Gaussian-averaged cubic vertex reduces exactly to the bare cubic BO derivative evaluated at the centroid position, \(\Phi^{(3)}=\phi^{(3)}\). The SSCHA result instead follows from optimizating the auxiliary Gaussian state on the complete BO potential \(V_{\rm BO}\). 
The consistent cubic vacuum-bubble correction is then
obtained by expanding around this optimized Gaussian reference and
adding $\Delta F_B[\Phi^{(3)}]$. To isolate the effect of this vertex dressing, we also evaluate the same vacuum diagram using the same SSCHA auxiliary propagator but replacing \(\Phi^{(3)}\) with the bare vertex \(\phi^{(3)}\). Although this latter curve and SCP\(+\Delta F_{\rm B}[\phi^{(3)}]\) employ the same bare cubic vertex, they remain distinct approximations because their Gaussian propagators are optimized on \(V_{\rm BO}\) and \(V_{\rm BO\leq 4}\), respectively.

\par The crossover from negative to positive thermal expansion is defined by \(\alpha(T^*)=0\), where \(\alpha=a_0^{-1}\partial a_0/\partial T\). The interpolated crossover temperatures are 433~K for SCP, 630~K for SCP\(+\Delta F_{\rm B}[\phi^{(3)}]\), 651~K for SSCHA, 672~K for SSCHA\(+\Delta F_{\rm B}[\Phi^{(3)}]\), and 753~K for SSCHA\(+\Delta F_{\rm B}[\phi^{(3)}]\). This progression makes the hierarchy of anharmonic effects quantitative. Replacing the quartic-truncated SCP reference with the Gaussian state optimized on the full BO potential raises \(T^*\) by 218~K. The cubic vacuum-bubble correction constructed
with the dressed vertices produces a further increase of 21~K,
yielding a total shift of 239~K relative to SCP. Thus, for the thermal-expansion crossover, the dominant change beyond quartic SCP originates from the full-potential Gaussian resummation, whereas the post-Gaussian cubic correction comes from the impact of the full BO potential on the auxiliary free energy itself, while the impact of the cubic correction is comparatively small. On the quartic-truncated reference obtained within SCP, the same cubic vacuum topology shifts \(T^*\) upward by 197~K, bringing SCP\(+\Delta F_{\rm B}[\phi^{(3)}]\) close to the uncorrected SSCHA result. The obtained SCP results here are consistent with previous calculations~\cite{oba2019first}.

\par Evaluating the correction with the bare vertex on the SSCHA propagator instead raises \(T^*\) by 102~K and moves the crossover toward the experimentally inferred onset of positive thermal expansion above approximately 1100~K~\cite{greve2010pronounced}. The much larger bare-vertex shift demonstrates the sensitivity of the free-energy equation of state to fluctuation-induced vertex renormalization. 
However, the bare-vertex result should be regarded as a diagnostic comparison, and its closer numerical proximity to experiment for this single thermodynamic observable does not by itself establish a more accurate beyond-SSCHA description.

\par To elucidate the microscopic origin of the negative thermal expansion, we analyze the mode-resolved Gr\"uneisen parameters of the optimized SSCHA Gaussian state, following the formulation derived in Sec.~\ref{app:sscha_gruneisen}. As shown in Fig.~\ref{fig:gruneisen}(a), the low-frequency phonon branches around the $M$-$R$ region exhibit strongly negative Gr\"uneisen parameters, consistent with the rotational and quasi-rotational motions of the ScF$_6$ octahedra. The cumulative frequency-resolved contributions in Fig.~\ref{fig:gruneisen}(b) shows that the negative thermal-expansion contribution is dominated by these low-frequency modes, whereas the higher-frequency optical phonons, particularly those around 60 meV, contribute positively. At 300 K, the negative contribution outweighs the positive one, resulting in a net negative thermal-expansion coefficient.

The temperature evolution of these contributions is shown in Fig.~\ref{fig:gruneisen}(c), evaluated at the equilibrium volumes obtained from the SSCHA free-energy equation of state. As the temperature increases, the negative contribution progressively weakens, while the positive contribution becomes stronger. The weakening of the negative contribution is consistent with the anharmonic hardening of the low-frequency soft modes and the reduction of their negative Gr\"uneisen parameters, whereas the growth of the positive contribution is associated with the
increasing heat-capacity weight of the higher-frequency phonons. Consequently, the total entropy volume derivative approaches zero around 600 K and becomes positive at higher temperatures, consistent with the negative to positive thermal expansion crossover obtained from the SSCHA free energy.

\begin{figure}[t]
    \centering
    \includegraphics[width=0.90\linewidth]{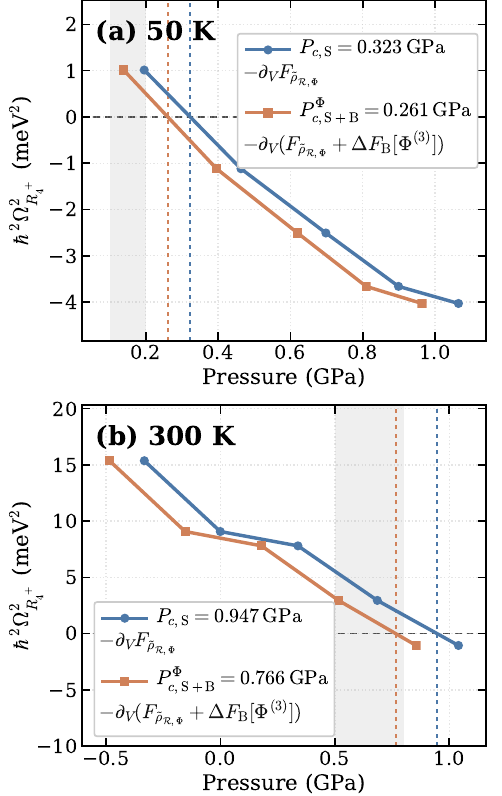}
    \caption{\label{fig:ScF} Lowest cubic-phase $R_4^+$ free-energy Hessian eigenvalue, $\Omega_{R_4^+}^{2}$, plotted against pressure at (a) 50~K and (b) 300~K. Blue curves use the SSCHA pressure $P_{\mathrm S}$, whereas orange curves use $P_{\mathrm{S+B}}^{\Phi}$. The horizontal dash-dotted line marks $\Omega_{R_4^+}^{2}=0$, the colored vertical dashed lines mark the corresponding calculated zero crossings, and the gray bands denote the experimental transition ranges~\cite{greve2010pronounced}. 
    }
\end{figure}

\subsection{Pressure–temperature local-stability line}
\label{sec:local-stability_line}
\par The pressure-induced cubic-to-rhombohedral transition in ScF$_3$ is well established experimentally~\cite{greve2010pronounced}. To illustrate how the structure evolves near this instability, we performed variable-cell SSCHA relaxations at 50 K, starting from an $R\bar{3}c$ structure constructed along the harmonic $R_4^+$ distortion. The relaxation procedure and the evolution of the octahedral tilt angle over successive, independently resampled SSCHA minimizations are described in Appendix~\ref{app:dp}.

\begin{figure}[t]
    \centering
    \includegraphics[width=\linewidth]{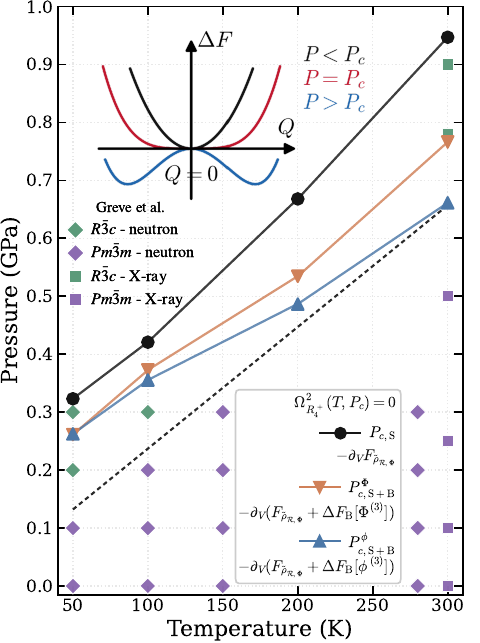}
    \caption{\label{fig:PB} Pressure--temperature local-stability line of cubic ScF$_3$. At each temperature, the same fixed-volume SSCHA Hessian data $\Omega_{R_4^{+}}^{2}(V_i,T)$ are remapped using $P_{\mathrm{S}}$, $P^{\Phi}_{\mathrm{S+B}}$, and $P^{\phi}_{\mathrm{S+B}}$, and the corresponding local-stability pressures $P_c(T)$ are determined from $\Omega_{R_4^{+}}^{2}(T,P_c)=0$. The inset schematically illustrates the evolution of the free-energy profile along the $R_4^{+}$ order parameter $Q$: the cubic configuration $Q=0$ is locally stable for $P<P_c$, marginal at $P=P_c$, and unstable for $P>P_c$. 
    Experimental data from Greve et al~\cite{greve2010pronounced} are shown as colored diamond and squares, while the black dashed line just represents the phase boundary.
    }
\end{figure}

\par Here, we instead quantify the cubic local-stability limit from the free-energy Hessian and examine how the cubic vacuum-bubble correction affects its volume-to-pressure mapping. Figs.~\ref{fig:ScF}(a) and \ref{fig:ScF}(b) show how the cubic vacuum-bubble free-energy correction modifies the pressure mapping of the cubic local-stability limit at 50 and 300~K. At each fixed volume \(V_i\), we calculate the lowest \(R_4^+\) eigenvalue of the SSCHA free-energy Hessian,
\begin{equation}
\left.
\Omega_{R_4^+}^{2}(V_i,T)
=
\frac{\partial^2 F_{ \tilde{\rho}_{\boldsymbol{\mathcal R},\mathbf{\Phi}}}}
{\partial Q_{R_4^+}^{2}}
\right|_{Q=0,V_i}.
\end{equation}
Over the same volume range, we fit \(F_{ \tilde{\rho}_{\boldsymbol{\mathcal R},\mathbf{\Phi}}}(V_i,T)\) and \(F_{ \tilde{\rho}_{\boldsymbol{\mathcal R},\mathbf{\Phi}}}(V_i,T)+\Delta F_{\mathrm B}[\Phi^{(3)}](V_i,T)\). Their volume derivatives define the SSCHA and the average-vertex-corrected  pressure mappings,
\(P_{\mathrm S}(V_i,T)=-\partial_VF_{ \tilde{\rho}_{\boldsymbol{\mathcal R},\mathbf{\Phi}}}(V_i,T)\) and
\(P_{\mathrm{S+B}}^{\Phi}(V_i,T)=-\partial_V({F_{ \tilde{\rho}_{\boldsymbol{\mathcal R},\mathbf{\Phi}}}+\Delta F_{\mathrm B}[\Phi^{(3)}])(V_i,T)}\), respectively. The same fixed-volume Hessian eigenvalue is assigned to either \(P_{\rm S}(V_i)\) or \(P_{\mathrm{S+B}}^{\Phi}(V_i)\), and a linear fit of \(\Omega_{R_4^+}^{2}\) against the corresponding pressure near \(\Omega_{R_4^+}^{2}=0\) yields the mapped local-stability pressure \(P_{c}(T)\). Importantly, $\Delta F_{\mathrm{B}}$ is included only in the equation of state used to map volume onto pressure; its curvature, $\partial^2\Delta F_{\mathrm{B}}/\partial Q_{R_4^{+}}^2$, is not included in the free-energy Hessian. The curves therefore represent pressure-remapped SSCHA local-stability limits, rather than local-stability limits obtained from a fully corrected free-energy Hessian or thermodynamic phase boundaries.
At 50 K, the uncorrected and corrected pressure mappings yield 0.323 and 0.261~GPa at 50~K, respectively; at 300 K, the corresponding values are 0.947 and 0.766~GPa.
Because the fixed-volume Hessian data are identical, these shifts arise entirely from the volume derivative of the cubic vacuum contribution. Thus, although \(\Delta F_{\mathrm B}\) is relatively small in magnitude, its effect on the equation of state can produce an appreciable displacement of the local-stability pressure near the soft-mode instability.

\par Fig.~\ref{fig:PB} extends this pressure-remapping analysis over the full temperature range by comparing the uncorrected SSCHA local-stability pressure $P_{c,\mathrm{S}}(T)$ with the two corrected mappings, $P^{\Phi}_{c,\mathrm{S+B}}(T)$ and $P^{\phi}_{c,\mathrm{S+B}}(T)$, on the same axes. In all three cases, the same fixed-volume SSCHA Hessian data $\Omega^{2}_{R_4^{+}}(V_i,T)$ are used. The corrected pressure mappings are defined as
$P^{X}_{\mathrm{S+B}}(V_i,T)
=-\partial_V\left(F_{\tilde{\rho}_{\mathbf{R},\Phi}}
+\Delta F_{\mathrm{B}}[X^{(3)}]\right)$,
with $X^{(3)}=\Phi^{(3)}$ or $\phi^{(3)}$. The two corrected calculations therefore share the same optimized SSCHA reference propagator $G^{\mathrm{S}}$ and the same cubic vacuum-bubble topology, differing only in the choice of the cubic  vertex for the correction. Their separation thus directly measures the effect of fluctuation dressing of the residual cubic interaction on the equation-of-state. The two corrections are nearly indistinguishable at low temperature but separate progressively upon heating. Both lower the mapped local-stability pressure relative to the uncorrected SSCHA result, with the bare vertex producing the larger shift and bringing the mapped local-stability line into closer numerical proximity to the experimental transition pressures. Within the systematic expansion around the optimized SSCHA Gaussian reference, however, the cubic vacuum-bubble correction is constructed from the Gaussian-averaged vertex $\Phi^{(3)}$, whereas the bare-vertex result is retained only as a diagnostic comparison. Its closer thermodynamic agreement therefore does not by itself establish a more accurate post-variational description, and the dynamical comparison below provides an independent test of the two vertex choices. 

\subsection{Soft-mode spectra and quasiparticle energies}

\par
\begin{figure}[t]
\centering
\includegraphics[width=\linewidth]{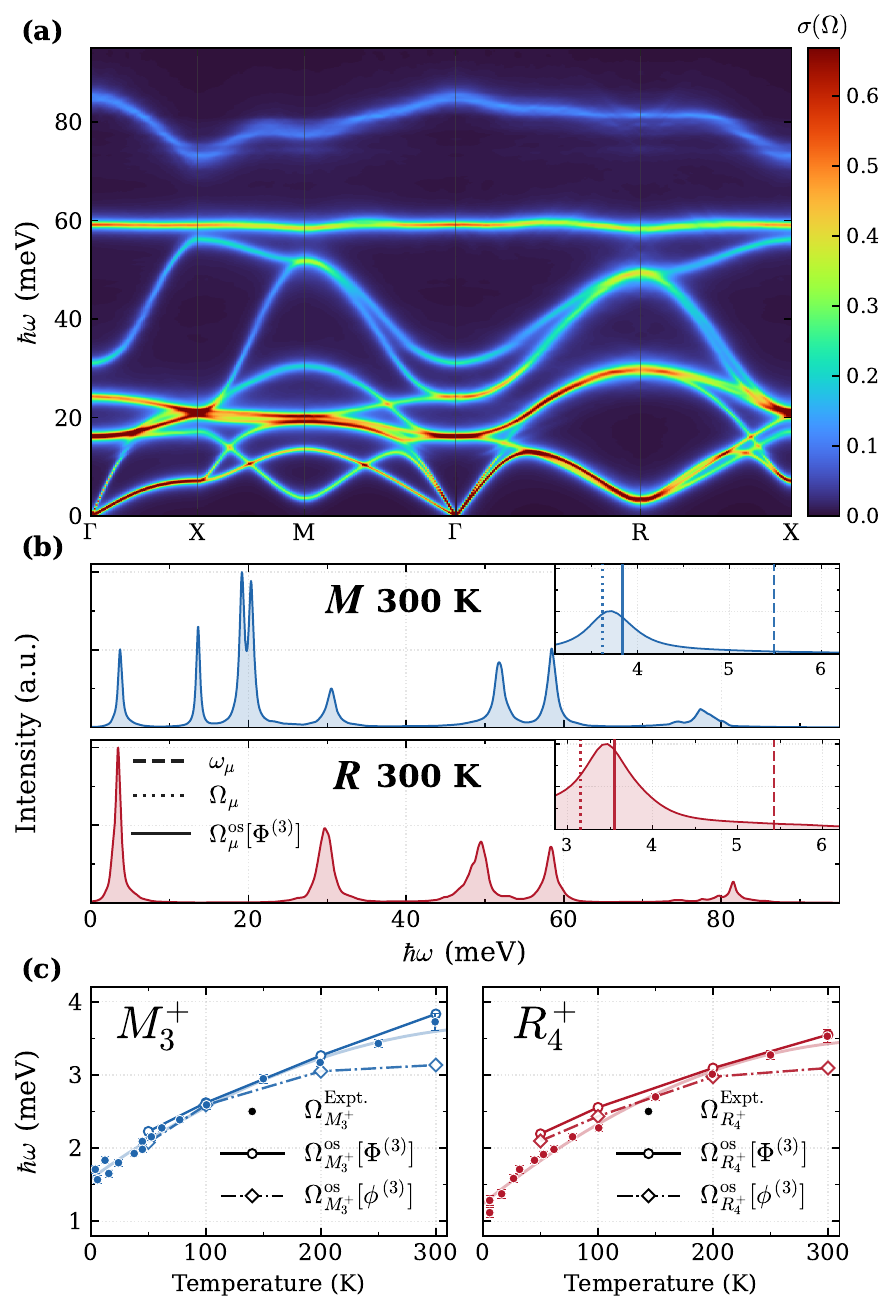}
\caption{\label{Fig5} Dynamical manifestation of fluctuation-dressed anharmonicity in cubic ScF$_3$. \textbf{(a)} Total spectral function $\sigma(\mathbf{q},\Omega)$ at 300~K along the high-symmetry path, calculated with the dynamical bubble self-energy constructed from $G^S$ and $\Phi^{(3)}$. \textbf{(b)} Spectral functions at the $M$ and $R$ points at 300~K. The insets magnify the low-energy $M_3^+$ and $R_4^+$ peaks. Vertical lines mark the auxiliary frequencies $\omega_\mu$ (dashed), free-energy-Hessian frequencies $\Omega_\mu$ (dotted), and quasiparticle energies $\Omega_\mu^{\mathrm{os}}[\Phi^{(3)}]$ (solid) within the one-shot approximation. \textbf{(c)} Temperature dependence of the $M_3^+$ and $R_4^+$
one-shot quasiparticle energies evaluated with the $\Phi^{(3)}$ and $\phi^{(3)}$ vertices. Experimental values are from Handunkanda~\textit{et al.}~\cite{handunkanda2015large}.}
\end{figure}

 We now examine the consequences of the average vertices on the dynamical phonon properties introduced in Sec.~\ref{sec:cubbic}. The frequency-resolved spectra are obtained from the Dyson equation built on the optimized SSCHA propagator, using the dynamical bubble self-energy \(\Pi_{ab}^{(\mathrm B)}[X](z)\). For \(X^{(3)}=\Phi^{(3)}\), this construction is precisely the bubble approximation to the TD-SCHA self-energy. The calculation with \(X^{(3)}=\phi^{(3)}\) uses the same reference propagator and dynamical procedure and serves only to assess the effect of replacing the fluctuation-dressed cubic vertex by its bare BO counterpart. Fig.~\ref{Fig5}(a) shows the total spectral function \(\sigma(\mathbf q,\Omega)\) along the high-symmetry path at \(300~\mathrm{K}\), obtained from the propagator containing the dynamical bubble self-energy. Most branches exhibit well-resolved quasiparticle peaks, while the low-energy octahedral-tilt modes at \(M\) and \(R\) remain finite-frequency excitations. 

\par Fig.~\ref{Fig5}(b) resolves the corresponding \(M_3^+\) and \(R_4^+\) peaks. The auxiliary frequency \(\omega_\mu\) characterizes the optimized Gaussian reference, whereas \(\Omega_\mu\), obtained from the static free-energy Hessian, characterizes the zero-frequency curvature of the free-energy surface. Neither quantity is defined by the pole of the finite-frequency propagator, and they therefore need not coincide with each other or with the maximum of the spectral peak. By contrast, the one-shot quasiparticle energy \(\Omega_\mu^{\mathrm{os}}[\Phi^{(3)}]\), obtained by incorporating the frequency-dependent cubic bubble through the one-shot quasiparticle equation~\cite{bianco2018high,monacelli2021time}, closely tracks the maxima of the full spectral function for both modes.

\par Fig.~\ref{Fig5}(c) then uses the one-shot energies to assess the role of fluctuation dressing in the dynamical cubic correction. The dressed and bare calculations use the same \(G^S\), the same two-phonon propagator \(\Lambda\), and the same one-shot quasiparticle equation. They differ only through the replacement \(\Phi^{(3)}\rightarrow\phi^{(3)}\) at the two cubic vertices. Up to \(200~\mathrm{K}\), the resulting \(M_3^+\) and \(R_4^+\) energies remain close, so that vertex dressing has only a modest dynamical effect in this temperature range. At \(300~\mathrm{K}\), however, the two predictions separate for both modes. The dressed-vertex results remain close to the measured energies~\cite{handunkanda2015large}, whereas the bare vertex systematically underestimates the energies of both
excitations. The increasing separation is consistent with the growing importance of the covariance contractions of fifth- and higher-order odd BO derivatives contained in \(\Phi^{(3)}\). Thus, although the bare-vertex vacuum correction shifts the thermodynamic observables toward experiment, the numerical improvement is not reproduced in the soft-mode dynamics and therefore suggests that the apparent thermodynamic improvement may result from error cancellation rather than a systematic improvement of the expansion. The dynamical measurements instead support the fluctuation-dressed vertex generated by the finite-temperature expansion around the optimized SSCHA state. Together with the thermodynamic results, this comparison supports
the reference-consistent dressed-vertex construction in
$\mathrm{ScF}_3$ and demonstrates that fluctuation-induced vertex
dressing is essential for quantitative predictions of the soft-mode
energies.

\section{CONCLUSIONS}
\label{sec:conclusions}
\par In summary, we have developed a finite-temperature post-variational expansion around the optimized SSCHA Gaussian reference. The cubic vacuum-bubble free-energy correction and its dynamical counterpart are generated by the same fluctuation-dressed cubic residual vertex, providing a unified description of non-Gaussian thermodynamics and lattice dynamics. Applied to scandium trifluoride, the full-potential Gaussian resummation accounts for the dominant change in the thermal-expansion crossover relative to quartic SCP, while the non-Gaussian correction produces a smaller but appreciable modification of the pressure-remapped local-stability limit. In the dynamical calculations, the fluctuation-dressed cubic interaction reproduces the measured soft-mode energies more closely, whereas the bare interaction oversoftens them at elevated temperature despite yielding closer thermodynamic agreement. This contrast suggests that the apparent thermodynamic improvement obtained with the bare vertex may arise from error cancellation and supports the dressed vertex derived from the optimized SSCHA Gaussian reference.

\begin{acknowledgments}
This work is supported by the Spanish Ministry of Science and Innovation, Grants No. PID2025-172166NB-C21 and PID2025-173011NA-I00, and the Severo Ochoa Centre of Excellence programme CEX2023-001286-S, all funded by MICIU/AEI/10.13039/501100011033.
The Department of Education, Universities and Research of the Eusko Jaurlaritza and the University of the Basque Country UPV/EHU (Grant No. 2122-26) is also acknowledged for financial support. 
Y.-W.F. thanks Prof. Lei Hu (Xi'an Jiaotong University) for sharing accounts of his experimental studies on the negative thermal expansion of ScF$_3$, which first drew the author's attention to this interesting phenomenon.
J.Y. acknowledges the China Scholarship Council (No. 202408330112). 
Technical and human support provided by DIPC Supercomputing Center is gratefully acknowledged.
\end{acknowledgments}

\appendix

\section{\MakeUppercase{Gaussian normal ordering and orthogonality}}
\label{app:normal_ordering}

We derive the Gaussian normal-ordering relations used in Sec.~\ref{sec:gaussian_normal_ordering}.

The generating function of the optimized Gaussian ionic density matrix is
\begin{align}
    \mathcal Z(\mathbf J)
    ={}&
    \frac{1}{\sqrt{\det(2\pi\boldsymbol{\Psi})}}
    \int d^N R
    \nonumber\\
    &\times
    \exp\left[
    -\frac12\mathbf u^{\mathrm T}
    \boldsymbol{\Psi}^{-1}\mathbf u
    +\mathbf J^{\mathrm T}\mathbf u
    \right].
    \label{eq:app_Z_integral}
\end{align}
Completing the square,
\begin{align}
    -\frac12\mathbf u^{\mathrm T}
    \boldsymbol{\Psi}^{-1}\mathbf u
    +\mathbf J^{\mathrm T}\mathbf u
    &=
    -\frac12
    (\mathbf u-\boldsymbol{\Psi}\mathbf J)^{\mathrm T}
    \boldsymbol{\Psi}^{-1}
    (\mathbf u-\boldsymbol{\Psi}\mathbf J)
    \nonumber\\
    &\quad+
    \frac12\mathbf J^{\mathrm T}
    \boldsymbol{\Psi}\mathbf J.
\end{align}
The change of variables
$\mathbf v=\mathbf u-\boldsymbol{\Psi}\mathbf J$
leaves a normalized Gaussian integral, giving
\begin{equation}
    \mathcal Z(\mathbf J)
    =
    e^{\frac12\mathbf J^{\mathrm T}
    \boldsymbol{\Psi}\mathbf J}.
    \label{eq:app_Z}
\end{equation}

The normal-ordering generating function is
\begin{equation}
    \mathcal N(\mathbf J,\mathbf u)
    =
    e^{\mathbf J^{\mathrm T}\mathbf u
    -\frac12\mathbf J^{\mathrm T}
    \boldsymbol{\Psi}\mathbf J}.
    \label{eq:app_N}
\end{equation}
Its derivatives generate the normal-ordered products.
For instance,
\begin{align}
    \frac{\partial\mathcal N}{\partial J_a}
    =
    \left(u_a-\sum_b\Psi_{ab}J_b\right)
    \mathcal N,
\end{align}
\begin{align}
    \left.
    \frac{\partial^2\mathcal N}
    {\partial J_a\partial J_b}
    \right|_{\mathbf J=\mathbf 0}
    =
    u_au_b-\Psi_{ab}.
\end{align}
Higher derivatives generate the corresponding Wick contractions.

By definition,
\begin{align}
    \left\langle
    \mathcal N(\mathbf J,\mathbf u)
    \right\rangle_{
    \tilde\rho_{\boldsymbol{\mathcal R},\boldsymbol\Phi}}
    =
    \frac{
    \left\langle e^{\mathbf J^{\mathrm T}\mathbf u}
    \right\rangle_{
    \tilde\rho_{\boldsymbol{\mathcal R},\boldsymbol\Phi}}
    }{\mathcal Z(\mathbf J)}
    =1.
    \label{eq:app_N_average}
\end{align}
Differentiating $n$ times with respect to $\mathbf J$ and setting $\mathbf J=\mathbf 0$
immediately gives
\begin{align}
    &\left\langle
    :u_{a_1}\cdots u_{a_n}:_{
    \tilde\rho_{\boldsymbol{\mathcal R},\boldsymbol\Phi}}
    \right\rangle_{
    \tilde\rho_{\boldsymbol{\mathcal R},\boldsymbol\Phi}}
    =0, n\geq1.
    \label{eq:app_zero_average}
\end{align}

To prove orthogonality, we introduce two independent source vectors $\mathbf J$ and
$\mathbf K$. Their generating functions satisfy
\begin{align}
    \left\langle
    \mathcal N(\mathbf J,\mathbf u)
    \mathcal N(\mathbf K,\mathbf u)
    \right\rangle_{
    \tilde\rho_{\boldsymbol{\mathcal R},\boldsymbol\Phi}}
    =
    \frac{\mathcal Z(\mathbf J+\mathbf K)}
    {\mathcal Z(\mathbf J)\mathcal Z(\mathbf K)}
    =
    \exp\left(
    \mathbf J^{\mathrm T}
    \boldsymbol{\Psi}\mathbf K
    \right).
    \label{eq:app_two_sources}
\end{align}
Expanding the exponential,
\begin{equation}
    e^{\mathbf J^{\mathrm T}\boldsymbol{\Psi}\mathbf K}
    =
    \sum_{r=0}^{\infty}
    \frac{1}{r!}
    \left(
    \sum_{ab}J_a\Psi_{ab}K_b
    \right)^r.
    \label{eq:app_cross_expansion}
\end{equation}
Each term contains equal powers of $\mathbf J$ and $\mathbf K$. Hence, differentiation of orders $n$ and $m$ with respect to the two sources gives zero at $\mathbf J=\mathbf K=\mathbf 0$ unless $n=m$.

For equal ranks, the derivatives give all possible pairings between the two products,
\begin{widetext}
\begin{equation}
    \Big\langle
    :u_{a_1}\cdots u_{a_n}:_{
    \tilde\rho_{\boldsymbol{\mathcal R},\boldsymbol\Phi}}
    \times
    :u_{b_1}\cdots u_{b_m}:_{
    \tilde\rho_{\boldsymbol{\mathcal R},\boldsymbol\Phi}}
    \Big\rangle_{
    \tilde\rho_{\boldsymbol{\mathcal R},\boldsymbol\Phi}}
    =
    \begin{cases}
    0, & n\neq m,\\[3pt]
    \displaystyle
    \sum_{\pi\in S_n}
    \prod_{i=1}^{n}\Psi_{a_i b_{\pi(i)}},
    & n=m,
    \end{cases}
    \label{eq:app_orthogonality}
\end{equation}
\end{widetext}
where $S_n$ denotes the permutations of $n$ elements.

Thus, different normal-ordered ranks are orthogonal, while equal-rank products retain contractions between the two factors.

\section{\MakeUppercase{Normal-ordered expansion of the BO potential}}
\label{app:BO_projection}
We derive the Gaussian-averaged vertices entering the normal-ordered expansion of the BO potential. For a positive-definite covariance matrix $\boldsymbol{\Psi}$,
the Gaussian normal-ordered polynomials form a complete orthogonal basis in the corresponding Gaussian $L^2$ space. The BO potential can therefore be expanded as
\begin{align}
    V_{\mathrm{BO}}
    ={}&
    C^{(0)}
    +\sum_{n=1}^{\infty}\frac{1}{n!}
    \sum_{a_1\cdots a_n}
    C^{(n)}_{a_1\cdots a_n}
    \nonumber\\
    &\times
    :u_{a_1}\cdots u_{a_n}:_{
    \tilde\rho_{\boldsymbol{\mathcal R},\boldsymbol\Phi}},
    \label{eq:app_BO_expansion}
\end{align}
where the tensors $C^{(n)}$ are symmetric.

Multiplying Eq.~\eqref{eq:app_BO_expansion} by a normal-ordered product of rank $m$ and taking the Gaussian average, the orthogonality relation derived in Appendix~\ref{app:normal_ordering} gives
\begin{align}
    &\left\langle
    V_{\mathrm{BO}}
    :u_{b_1}\cdots u_{b_m}:_{
    \tilde\rho_{\boldsymbol{\mathcal R},\boldsymbol\Phi}}
    \right\rangle_{
    \tilde\rho_{\boldsymbol{\mathcal R},\boldsymbol\Phi}}
    \nonumber\\
    &=
    \frac{1}{m!}
    \sum_{a_1\cdots a_m}
    C^{(m)}_{a_1\cdots a_m}
    \sum_{\pi\in S_m}
    \prod_{i=1}^{m}
    \Psi_{a_i b_{\pi(i)}}.
    \label{eq:app_BO_projection_1}
\end{align}
Since $C^{(m)}$ is symmetric, all $m!$
permutations give identical contributions.
Thus,
\begin{align}
    &\left\langle
    V_{\mathrm{BO}}
    :u_{b_1}\cdots u_{b_m}:_{
    \tilde\rho_{\boldsymbol{\mathcal R},\boldsymbol\Phi}}
    \right\rangle_{
    \tilde\rho_{\boldsymbol{\mathcal R},\boldsymbol\Phi}}
    \nonumber\\
    &=
    \sum_{a_1\cdots a_m}
    \left(
    \prod_{i=1}^{m}\Psi_{b_i a_i}
    \right)
    C^{(m)}_{a_1\cdots a_m}.
    \label{eq:app_BO_projection_2}
\end{align}

The multivariate Gaussian integration-by-parts identity reads~\cite{bianco2017second}
\begin{align}
    \left\langle
    \frac{\partial \mathds{O}}{\partial u_a}
    \right\rangle_{
    \tilde\rho_{\boldsymbol{\mathcal R},\boldsymbol\Phi}}
    =
    \sum_b(\Psi^{-1})_{ab}
    \left\langle
    u_b \mathds{O}
    \right\rangle_{
    \tilde\rho_{\boldsymbol{\mathcal R},\boldsymbol\Phi}}.
    \label{eq:app_BO_IBP}
\end{align}
This is Eq.~(C1) of Ref.~\cite{bianco2017second}. Applying it successively to the normal-ordering generating function
\begin{equation}
    \mathcal N(\mathbf J,\mathbf u)
    =
    \exp\left[
    \mathbf J^{\mathrm T}\mathbf u
    -\frac12\mathbf J^{\mathrm T}
    \boldsymbol{\Psi}\mathbf J
    \right],
\end{equation}
we obtain
\begin{align}
    \frac{\partial}{\partial J_b}
    \left\langle
    V_{\mathrm{BO}}\mathcal N
    \right\rangle_{
    \tilde\rho_{\boldsymbol{\mathcal R},\boldsymbol\Phi}}
    =
    \sum_a\Psi_{ba}
    \left\langle
    \frac{\partial V_{\mathrm{BO}}}{\partial u_a}
    \mathcal N
    \right\rangle_{
    \tilde\rho_{\boldsymbol{\mathcal R},\boldsymbol\Phi}}.
    \label{eq:app_BO_IBP_gen}
\end{align}
Repeated differentiation with respect to the
source components, followed by evaluation at
$\mathbf J=\mathbf 0$, yields
\begin{align}
    &\left\langle
    V_{\mathrm{BO}}
    :u_{b_1}\cdots u_{b_m}:_{
    \tilde\rho_{\boldsymbol{\mathcal R},\boldsymbol\Phi}}
    \right\rangle_{
    \tilde\rho_{\boldsymbol{\mathcal R},\boldsymbol\Phi}}
    \nonumber\\
    &=
    \sum_{a_1\cdots a_m}
    \left(
    \prod_{i=1}^{m}\Psi_{b_i a_i}
    \right)
    \left\langle
    \frac{\partial^m V_{\mathrm{BO}}}
    {\partial u_{a_1}\cdots\partial u_{a_m}}
    \right\rangle_{
    \tilde\rho_{\boldsymbol{\mathcal R},\boldsymbol\Phi}}.
    \label{eq:app_BO_IBP_result}
\end{align}

Comparing Eqs.~\eqref{eq:app_BO_projection_2}
and~\eqref{eq:app_BO_IBP_result}, and using
the permutation symmetry of $\boldsymbol{\Psi}$, gives
\begin{equation}
    C^{(m)}_{a_1\cdots a_m}
    =
    \left\langle
    \frac{\partial^m V_{\mathrm{BO}}}
    {\partial u_{a_1}\cdots\partial u_{a_m}}
    \right\rangle_{
    \tilde\rho_{\boldsymbol{\mathcal R},\boldsymbol\Phi}}
    \equiv
    \Phi^{(m)}_{a_1\cdots a_m}.
    \label{eq:app_BO_coefficients}
\end{equation}
Equivalently, the coefficients can be evaluated
directly from the Gaussian projection,
\begin{align}
    \Phi^{(m)}_{a_1\cdots a_m}
    ={}&
    \sum_{b_1\cdots b_m}
    \left[
    \prod_{i=1}^{m}
    (\Psi^{-1})_{a_i b_i}
    \right]
    \nonumber\\
    &\times
    \left\langle
    V_{\mathrm{BO}}
    :u_{b_1}\cdots u_{b_m}:_{
    \tilde\rho_{\boldsymbol{\mathcal R},\boldsymbol\Phi}}
    \right\rangle_{
    \tilde\rho_{\boldsymbol{\mathcal R},\boldsymbol\Phi}}.
    \label{eq:app_BO_coefficients_projection}
\end{align}
The zeroth-order coefficient follows from the
vanishing average of all positive-rank
normal-ordered products,
\begin{equation}
    C^{(0)}
    =
    \langle V_{\mathrm{BO}}\rangle_{
    \tilde\rho_{\boldsymbol{\mathcal R},\boldsymbol\Phi}}.
\end{equation}

\section{\MakeUppercase{Derivation of the cubic vacuum-bubble
free-energy correction}}
\label{app:cubic_bubble}

We provide the Wick contractions and
imaginary-time integration leading to the
cubic vacuum-bubble correction in Sec.~\ref{sec:cubic_bubble}.

Since the cubic residual vertices are normal ordered
with respect to the optimized Gaussian density matrix
$\tilde{\rho}_{\boldsymbol{\mathcal R},\mathbf{\Phi}}$,
all intra-vertex contractions vanish.
Wick's theorem therefore gives
\begin{equation}
\begin{aligned}
&\Big\langle T_\tau
:u_a u_b u_c:(\tau)
:u_d u_e u_f:(\tau')
\Big\rangle_{
\tilde{\rho}_{\boldsymbol{\mathcal R},\mathbf{\Phi}}}
\\
&\quad=
\sum_{\pi\in S_3}
G^S_{a j_{\pi(1)}}(\tau-\tau')
G^S_{b j_{\pi(2)}}(\tau-\tau')
G^S_{c j_{\pi(3)}}(\tau-\tau'),
\end{aligned}
\label{eq:app_cubic_wick}
\end{equation}
where $(j_1,j_2,j_3)=(d,e,f)$ and $S_3$
denotes the permutation group of three indices.

Because the cubic vertices are fully symmetric,
all $3!$ contractions contribute equally after
summation over Cartesian indices. The combinatorial
factor is therefore $3!/[2(3!)^2]=1/12$, giving
\begin{equation}
\begin{aligned}
\Delta F_{\mathrm B}
={}&-\frac{1}{12\beta}
\sum_{abcdef}
\Phi_{abc}^{(3)}\Phi_{def}^{(3)}
\int_0^\beta d\tau
\int_0^\beta d\tau'
\\
&\quad\times
G^S_{ad}(\tau-\tau')
G^S_{be}(\tau-\tau')
G^S_{cf}(\tau-\tau').
\end{aligned}
\label{eq:app_cubic_cartesian}
\end{equation}

Using imaginary-time translational invariance
and the bosonic periodicity of the propagators,
\begin{equation}
\int_0^\beta d\tau
\int_0^\beta d\tau'\,
f(\tau-\tau')
=
\beta\int_0^\beta d\tau\,f(\tau).
\label{eq:app_cubic_translation}
\end{equation}
The Cartesian propagator can be expanded in
the auxiliary normal-mode basis as
\begin{equation}
G^S_{ab}(\tau)
=
\frac{1}{\sqrt{M_aM_b}}
\sum_\mu
e_a^\mu e_b^{\mu *}G_\mu^S(\tau).
\label{eq:app_cubic_mode_G}
\end{equation}

Substituting this expression into
Eq.~\eqref{eq:app_cubic_cartesian}, and using the
mass-weighted cubic vertex defined in the main text,
we obtain
\begin{equation}
\begin{aligned}
\Delta F_{\mathrm B}
&=
-\frac{1}{12}
\sum_{\mu\nu\lambda}
|D_{\mu\nu\lambda}^{(3)}|^2
I_{\mu\nu\lambda},
\\
I_{\mu\nu\lambda}
&=
\int_0^\beta d\tau\,
G_\mu^S(\tau)
G_\nu^S(\tau)
G_\lambda^S(\tau).
\end{aligned}
\label{eq:app_cubic_mode_integral}
\end{equation}

For $0<\tau<\beta$, inserting the auxiliary
harmonic propagators gives
\begin{equation}
\begin{aligned}
I_{\mu\nu\lambda}
={}&
\frac{1}{8\omega_\mu\omega_\nu\omega_\lambda}
\int_0^\beta d\tau
\\
&\times
\prod_{\alpha=\mu,\nu,\lambda}
\left[
(1+n_\alpha)e^{-\omega_\alpha\tau}
+n_\alpha e^{\omega_\alpha\tau}
\right].
\end{aligned}
\label{eq:app_cubic_product}
\end{equation}

The product contains $2^3=8$ exponential terms.
Using
$(1+n_\alpha)e^{-\beta\omega_\alpha}=n_\alpha$,
terms related by $\tau\rightarrow\beta-\tau$
can be paired. For example, defining
$\Omega=\omega_\mu+\omega_\nu+\omega_\lambda$,
the pair containing three positive or three
negative exponents satisfies
\begin{equation}
\begin{aligned}
&\int_0^\beta d\tau\,
\Big[
(1+n_\mu)(1+n_\nu)(1+n_\lambda)
e^{-\Omega\tau}
\\
&\qquad\qquad+
n_\mu n_\nu n_\lambda e^{\Omega\tau}
\Big]
\\
&=
2\frac{
(1+n_\mu)(1+n_\nu)(1+n_\lambda)
-n_\mu n_\nu n_\lambda
}{\Omega}.
\end{aligned}
\label{eq:app_cubic_pair}
\end{equation}

The remaining three pairs are obtained in the same
way. The complete imaginary-time integral is
therefore

\begin{equation}
\begin{aligned}
I_{\mu\nu\lambda}
=
\frac{1}{4\omega_\mu\omega_\nu\omega_\lambda}
\Bigg[
&
\frac{
(1+n_\mu)(1+n_\nu)(1+n_\lambda)
-n_\mu n_\nu n_\lambda
}{
\omega_\mu+\omega_\nu+\omega_\lambda
}\\
&+
\frac{
n_\mu(1+n_\nu+n_\lambda)-n_\nu n_\lambda
}{
-\omega_\mu+\omega_\nu+\omega_\lambda
}\\
&+
\frac{
n_\nu(1+n_\mu+n_\lambda)-n_\mu n_\lambda
}{
\omega_\mu-\omega_\nu+\omega_\lambda
}
\\
&+
\frac{
n_\lambda(1+n_\mu+n_\nu)-n_\mu n_\nu
}{
\omega_\mu+\omega_\nu-\omega_\lambda
}
\Bigg].
\end{aligned}
\label{eq:app_cubic_unsymmetrized}
\end{equation}

Eq.~\eqref{eq:app_cubic_unsymmetrized} is the
unsymmetrized result. When the mode summation
includes all ordered triplets, the last three
terms give identical contributions after
permutation of the dummy indices. Hence, the
three difference-frequency terms combine into
a single contribution multiplied by three.
Substituting into Eq.~\eqref{eq:app_cubic_mode_integral}
gives the compact expression in
Eq.~\eqref{eq:cubic_vacuum_final}, with the
overall coefficient $-1/(12\times4)=-1/48$.

Finally, the apparent singularities associated
with vanishing difference-frequency denominators
are removable. For example, when
$\omega_\mu\rightarrow\omega_\nu+\omega_\lambda$,
the numerator vanishes simultaneously and
\begin{equation}
\begin{aligned}
&\lim_{\omega_\mu\to\omega_\nu+\omega_\lambda}
\frac{
n_\mu(1+n_\nu+n_\lambda)-n_\nu n_\lambda
}{
-\omega_\mu+\omega_\nu+\omega_\lambda
}
\\
&\qquad=
\beta(1+n_\mu)n_\nu n_\lambda.
\end{aligned}
\label{eq:app_cubic_resonance}
\end{equation}

The other difference-frequency terms have
analogous finite limits. Therefore, the
imaginary-time integral remains finite at
these resonances.

\section{\MakeUppercase{DEEP POTENTIAL VALIDATION AND PRESSURE-DEPENDENT STRUCTURAL RELAXATIONS}}
\label{app:dp}
\begin{figure}[t]
    \centering
    \includegraphics[width=\linewidth]{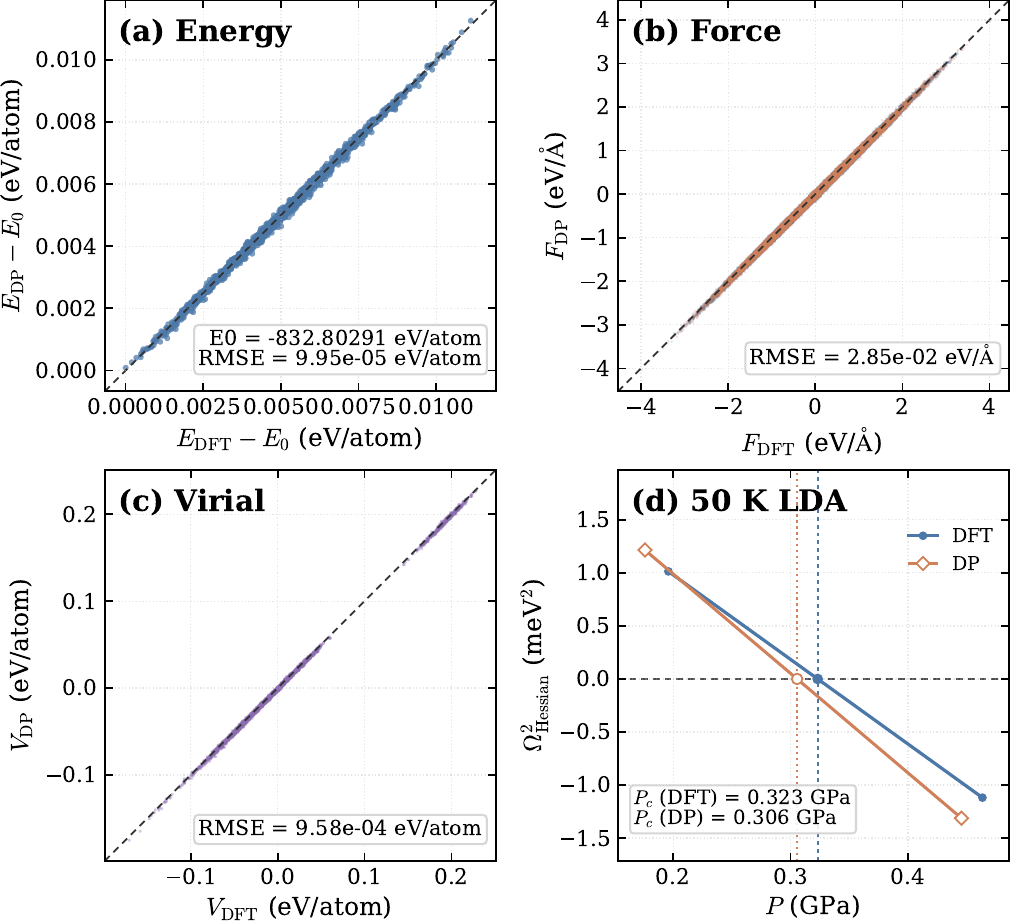}
    \caption{\label{fig:dp_validation} Validation of the Deep Potential (DP) model against DFT. (a)-(c) Comparison of the energies, atomic forces, and virial stresses, with dashed lines indicating perfect agreement. (d) Pressure dependence of the lowest cubic-phase $R_4^+$ free-energy-Hessian eigenvalue at 50 K. The vertical dotted lines mark the critical pressures obtained from DFT (blue line) and DP (orange line).
    }
\end{figure}

\par It is experimentally established that ScF$_3$ undergoes a pressure-induced transition from the cubic $Pm\bar{3} m$ phase to the rhombohedral $R\bar{3}c$ phase~\cite{greve2010pronounced}. As a starting point for studying this instability, we build an initial $R\bar{3}c$ structure by displacing the atoms along the unstable harmonic $R_4^+$ phonon modes at $\mathbf{q} _R=(0.5, 0.5, 0.5)$. The distortion is parameterized by a three-component order parameter \(Q _{R_4^+}=(Q_1,Q_2,Q_3),\) with atomic displacements 
\begin{equation}
\Delta u _{a,l}=\frac{1}{\sqrt{M_a} } \sum_{j =1} ^{3}Q_{j }e^a_{j}(\mathbf{q} _R)e^{i\mathbf{q} _R\cdot \mathbf{T}_l },
\end{equation}
where $\Delta u _{a,l}$ is the displacement associated with degree of freedom 
$a$ in cell $l$, $M_a$ is the mass of atom $a$, $e^a_{j}(\mathbf{q} _R)$ denotes the polarization vector of $R_4^+$ mode, and $\mathbf{T}_l$ is the lattice translation vector. Along the equal-amplitude direction $(Q,Q,Q)$, this gives the $R\bar{3}c$ distortion with the $a^-a^-a^-$~\cite{LWN-distortion2017PRB} octahedral-tilt pattern. As $Q\rightarrow 0$, the cubic structure is recovered continuously.

\begin{figure}[b]
    \centering
    \includegraphics[width=\linewidth]{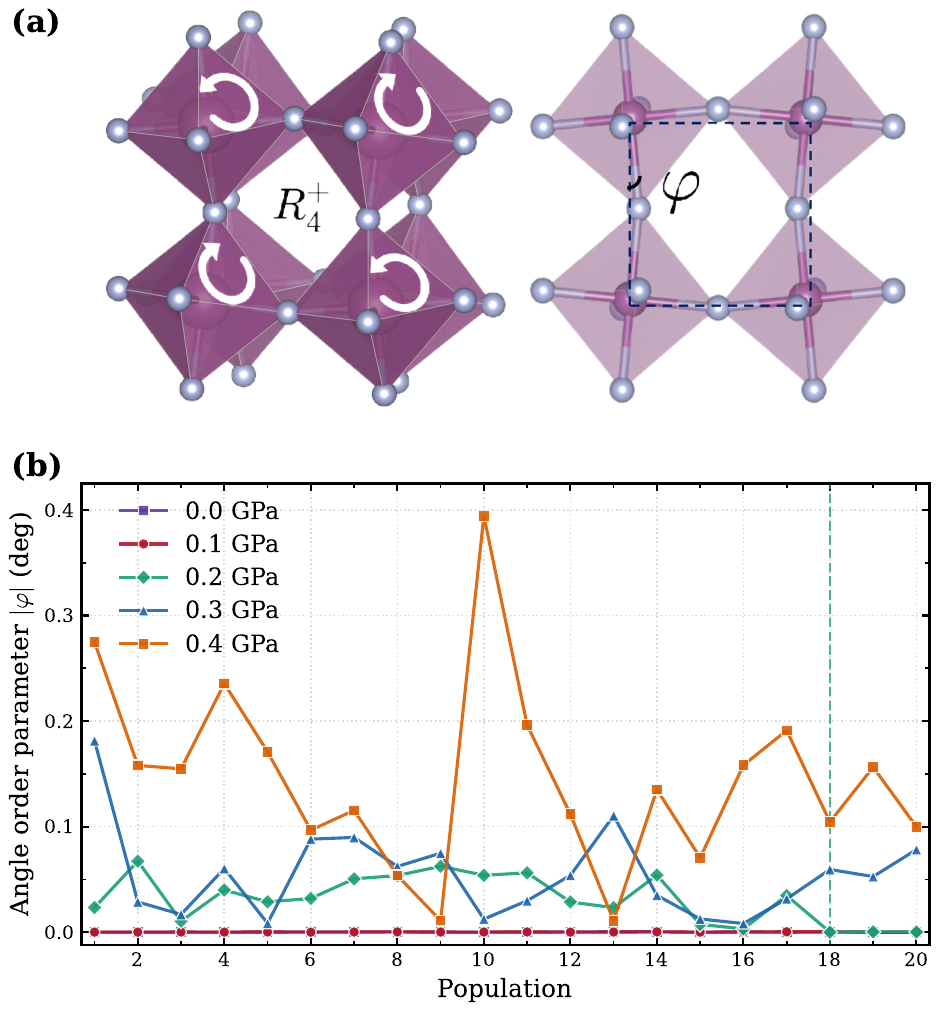}
    \caption{\label{fig:angle} 
    Pressure-dependent structural relaxation of ScF$_3$ at 50 K. (a) Schematic illustration of the \(R_4^+\) octahedral-tilt pattern (left) and definition of the octahedral tilt angle \(\varphi\) (right). (b) Absolute octahedral tilt angle $|\varphi|$ from 20 successive variable-cell SSCHA minimizations at several external pressures. Each marker is the optimized centroid configuration from one independently resampled SSCHA minimization, initialized from the previously optimized structure. The relaxations are constrained to the $R\bar{3}c$ symmetry subspace, which continuously includes the cubic configuration at $\varphi=0$. The green vertical dash-dotted line marks the population at which the 0.2 GPa relaxation reaches zero tilt and stays there through all later minimizations.
    }
\end{figure}

\par The SSCHA structural relaxations in the $R\bar{3}c$ symmetry subspace were accelerated using a Deep Potential (DP) model trained with DeePMD-kit~\cite{wang2018deepmd,zeng2023deepmd}. The training configurations were sampled from the relevant SSCHA distributions and labeled by DFT calculations of their energies, atomic forces, and virial stresses. As shown in Fig.~\ref{fig:dp_validation}(a-c), the model yields root-mean-square errors of
$9.95\times10^{-5}$ eV/atom, $2.85\times10^{-2}$ eV/\AA, and $9.58\times10^{-4}$ eV/atom for the energies, forces, and virials, respectively. To further assess the reliability of the DP model for the structural instability, Fig.~\ref{fig:dp_validation}(d) compares the pressure dependence of the lowest cubic-phase $R_4^+$ free-energy-Hessian eigenvalue at 50 K obtained from DFT and DP. The two calculations exhibit nearly the same pressure-induced softening, yielding local-stability pressures of 0.323 and 0.306 GPa, respectively. The DP model was employed for the symmetry-broken structural relaxations, while the cubic-phase free-energy Hessian used in the main analysis was evaluated directly from DFT.

\par Starting from the distorted $R\bar{3}c$ structure, we carried out variable-cell SSCHA structural relaxations at 50 K for several external pressures. Both the centroid positions and the lattice parameters were optimized under $R\bar{3}c$ symmetry constraints until the SSCHA stress matched the imposed hydrostatic pressure. At each pressure, we performed 20 consecutive SSCHA minimizations with 8000 configurations per population. Each minimization was fully converged and independently resampled, starting from the previously optimized structure. The resulting octahedral tilt angles $\varphi$
are shown in Fig.~\ref{fig:angle}.

\par At 0.2 GPa, the optimized tilt angle eventually falls to zero and stays there through all subsequent independently resampled minimizations. At 0.3 GPa, close to the cubic local-stability limit, the optimized tilt angle varies considerably from one population to the next, switching between nearly cubic and finite-tilt configurations. At 0.4 GPa, the near-cubic configuration is no longer
consistently retained. This behavior reflects the softening of the free-energy curvature along the $R_4^+$ direction near the structural instability, which makes the stochastic minimization more sensitive to statistical fluctuations.
These relaxations give a qualitative picture of the approaching cubic instability, but by themselves they neither determine the critical pressure nor establish that multiple free-energy minima exist. The cubic local-stability limit is instead obtained quantitatively from the SSCHA free-energy Hessian, as discussed in Sec.~\ref{sec:local-stability_line}.

\section{\MakeUppercase{Convergence of the free-Energy equation of State}}
\label{app:qmesh_convergence}
\begin{figure}[b]
    \centering
    \includegraphics[width=\linewidth]{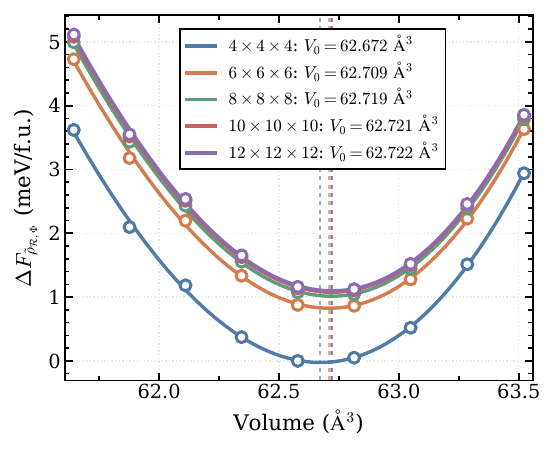}
    \caption{\label{fig:EOS} Convergence of the SSCHA free-energy equation of state at 50~K with respect to the reciprocal-space interpolation mesh. The auxiliary force constants were Fourier interpolated onto uniform $N_q\times N_q\times N_q$ meshes, and the resulting frequencies were used to evaluate the mode sum in Eq.~\eqref{H_F}. Open circles denote the calculated free energies, solid curves are the corresponding EOS fits, and vertical dashed lines indicate the equilibrium volumes $V_0$ obtained from their minima. The $8^3$, $10^3$, and $12^3$ meshes yield nearly indistinguishable curves and equilibrium volumes, demonstrating convergence with respect to the interpolation mesh.
    }
\end{figure}

\par The reciprocal-space convergence of the mode sum entering the SSCHA variational free energy was assessed by Fourier interpolating the optimized auxiliary force constants onto uniform $N_q\times N_q\times N_q$ meshes. To construct the free-energy
equation of state, the cubic lattice parameter was isotropically scaled according to
\begin{equation}
    \eta=\frac{a-a_{\mathrm{ref}}}{a_{\mathrm{ref}}},
\end{equation}
with $\eta$ ranging from $-0.5\%$ to $+0.5\%$ in increments of $0.125\%$. $a_{\rm ref}$ corresponds to the undeformed lattice constant. This procedure yielded nine lattice parameters and their corresponding volumes.

\par Fig.~\ref{fig:EOS} shows the resulting free-energy equations of state at 50~K. The $4^3$ interpolation mesh is clearly insufficient, producing a visibly different free-energy curve and equilibrium volume. By contrast, the free-energy equations of state (EOS), derived from a Birch-Murnaghan fit, and their corresponding equilibrium volumes are both effectively converged for meshes of $8^3$ and denser. In particular, the equilibrium volume changes by only $0.001$~\AA$^3$ between the $10^3$ and $12^3$ meshes. A $12\times12\times12$ reciprocal-space interpolation mesh was therefore adopted for the evaluation of the SSCHA variational free energy in all subsequent calculations.

\section{\MakeUppercase{Bubble approximation in the free-energy Hessian}}
\label{app:bubble}
 \begin{figure}[t]
    \centering
    \includegraphics[width=\linewidth]{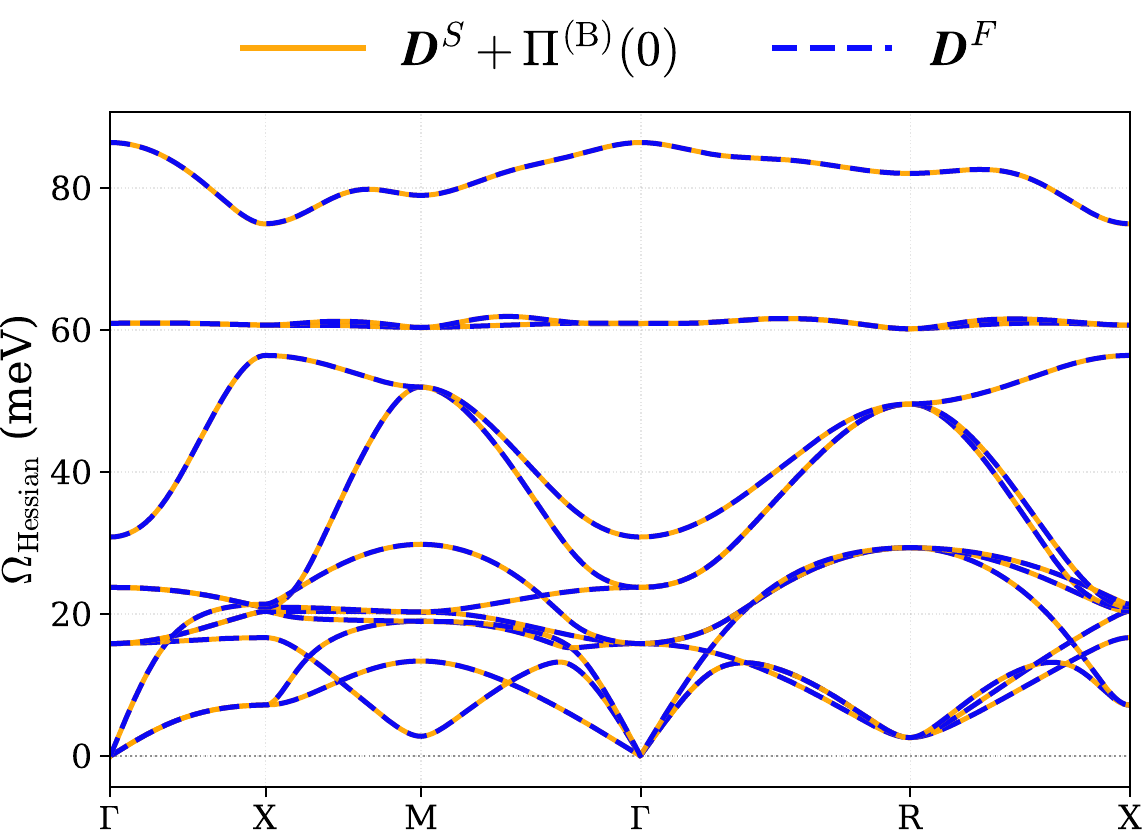}
    \caption{\label{fig:Hessian_frequency} Comparison of the full and bubble-approximated SSCHA free-energy Hessians for the cubic ScF$_3$ at 50~K. Free-energy Hessian frequencies along a high-symmetry path obtained from the full expression $D^{F}$ (blue dashed lines) and the static bubble approximation $D^{S}+\boldsymbol{\Pi}^{(\mathrm{B})}(0)$ (orange solid lines). 
    }
\end{figure}

\par As shown in Fig.~\ref{fig:Hessian_frequency}, the free-energy Hessian spectra of cubic ScF$_3$ obtained from the full SSCHA expression, Eq.~(\ref{Full}), and from the static bubble approximation, Eq.~(\ref{Bubble}), are nearly indistinguishable along the entire high-symmetry path at 50~K. In particular, no appreciable difference is visible for the soft $R_4^+$ mode at the $R$ point. This indicates that the contributions retained in the full Hessian beyond the static bubble approximation are negligible under these conditions, and that the dominant anharmonic renormalization of the positional free-energy Hessian is already captured at the bubble level.

\section{\MakeUppercase{Volume interpolation of the cubic vacuum-bubble free-energy correction}}
\label{app:bubble_volume_fit}
\begin{figure}[b]
    \centering
    \includegraphics[width=\linewidth]{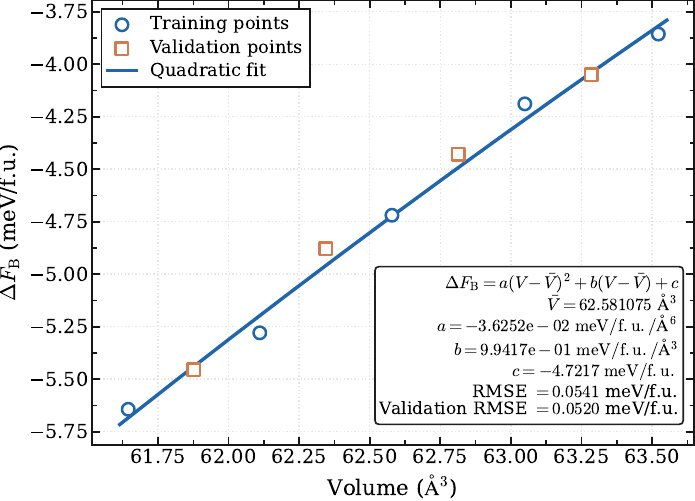}
    \caption{\label{fig:bubble_term} Volume dependence of the cubic vacuum-bubble free-energy correction
$\Delta F_B$ at 300~K. Blue circles and orange squares denote the training and validation points, respectively, and the solid line is the centered quadratic interpolation $\Delta F_{\rm B}(V)=a(V-\bar V)^2+b(V-\bar V)+c$. The comparable training and validation root-mean-square errors confirm the stability of the interpolation over the sampled volume range.
}
\end{figure}

\par The cubic vacuum-bubble free-energy correction $\Delta F_{\rm B}$ was evaluated at a discrete set of volumes and
interpolated independently at each temperature using a centered
quadratic polynomial,
\begin{equation}
    \Delta F_{\rm B}(V)
    =
    a(V-\bar V)^2+b(V-\bar V)+c,
\end{equation}
where \(\bar V)\) is a reference volume chosen near the center of the sampled interval, and \(a\), \(b\), and \(c\) are temperature-dependent fitting coefficients with units of energy per volume squared, energy per volume, and energy, respectively. This interpolation provides a smooth representation of the volume dependence of $\Delta F_{\rm B}$ for its subsequent inclusion in the corrected free-energy curves. To assess the robustness of the interpolation, the calculated data were divided into training and validation sets. As shown in Fig.~\ref{fig:bubble_term}, the validation points closely follow the quadratic curve obtained from the training data. The comparable training and validation root-mean-square errors further indicate that the fitted form accurately represents $\Delta F_{\rm B}(V)$ over the sampled volume range without appreciable overfitting.

\bibliography{ref}

\end{document}